\documentclass[manuscript,screen,nonacm]{acmart}
\ifdefined\pdfobjcompresslevel\pdfobjcompresslevel=0\fi

\usepackage{float}
\usepackage{placeins}
\usepackage{multirow}
\usepackage{amsmath,amsfonts}

\usepackage{amssymb}
\usepackage{amsthm}
\usepackage{siunitx}
\usepackage{algorithm}
\usepackage{algpseudocode}
\usepackage{array}
\usepackage{tabularx}
\usepackage{booktabs}
\usepackage{graphicx}
\usepackage{listings}
\usepackage{xcolor}
\usepackage[most]{tcolorbox}
\usepackage{threeparttable}
\usepackage{enumitem}
\usepackage{url}
\usepackage{tikz}
\usetikzlibrary{positioning,arrows.meta,fit}

\makeatletter
\renewcommand\paragraph{\@startsection{paragraph}{4}{\parindent}%
  {-.5\baselineskip \@plus -2\p@ \@minus -.2\p@}%
  {-3.5\p@}%
  {\ACM@NRadjust\@parfont}}
\renewcommand{\@secfont}{\sffamily\bfseries\section@raggedright}
\makeatother

\newcolumntype{L}[1]{>{\raggedright\arraybackslash}p{#1}}
\newcolumntype{C}[1]{>{\centering\arraybackslash}p{#1}}
\newcolumntype{Y}{>{\raggedright\arraybackslash}X}
\newcommand{\deiatablenote}[1]{%
  \par\vspace{2pt}%
  \noindent\begin{minipage}{\linewidth}%
    \footnotesize\raggedright\setlength{\parindent}{0pt}%
    \textit{Note.}~#1%
  \end{minipage}\par
}

\usepackage{etoolbox}
\AtBeginEnvironment{table}{\small\setlength{\tabcolsep}{6pt}}
\AtBeginEnvironment{table*}{\small\setlength{\tabcolsep}{6pt}}

\newtcolorbox{rqanswer}[1]{%
  enhanced,
  colback=white,
  colframe=black!65,
  colbacktitle=black!75,
  coltitle=white,
  fonttitle=\bfseries,
  title={#1},
  boxrule=0.4pt,
  arc=1mm,
  outer arc=1mm,
  titlerule=0pt,
  left=5pt,
  right=5pt,
  top=4pt,
  bottom=4pt,
  before skip=8pt,
  after skip=8pt,
}

\newtcolorbox{implicationbox}{
  enhanced,
  breakable,
  colback=black!6,
  colframe=black!55,
  boxrule=0.45pt,
  arc=2.5mm,
  left=7pt,
  right=7pt,
  top=6pt,
  bottom=6pt,
  before skip=6pt,
  after skip=8pt,
  drop shadow={black!25!white},
}

\lstdefinelanguage{ArkTS}{
  morekeywords={struct,Entry,Component,State,Prop,Link,Builder,build,aboutToAppear,
    async,await,let,const,this,return,if,else,for,of,new,throw,try,catch},
  sensitive=true,
  morecomment=[l]{//},
  morecomment=[s]{/*}{*/},
  morestring=[b]',
  morestring=[b]",
  morestring=[b]`
}

\setcopyright{none}
\begin{document}

\title[Evidence-Grounded Crash Repair for ArkTS]{%
DEIA: Evidence-Grounded Crash Repair with Change-Aware Test Design for HarmonyOS ArkTS}

\author{Xuhui Lu}
\affiliation{%
  \institution{Beihang University}
  \city{Beijing}
  \country{China}}
\email{luxuhui@buaa.edu.cn}

\author{Zhihao Liu}
\affiliation{%
  \institution{Beihang University}
  \city{Beijing}
  \country{China}}
\email{lzh123698745@buaa.edu.cn}

\author{Xiaoke Fang}
\affiliation{%
  \institution{Beihang University}
  \city{Beijing}
  \country{China}}
\email{zy2403505@buaa.edu.cn}

\author{Yuhuan Fan}
\affiliation{%
  \institution{Beihang University}
  \city{Beijing}
  \country{China}}
\email{zb2606301@buaa.edu.cn}

\author{Yanzhe Tang}
\affiliation{%
  \institution{Beihang University}
  \city{Beijing}
  \country{China}}
\email{zy2503117@buaa.edu.cn}

\author{Xiaohui Wan}
\affiliation{%
  \institution{Suzhou Aerospace Information Research Institute}
  \city{Suzhou}
  \country{China}}
\email{wanxh@aircas.ac.cn}

\author{Xiaoyu Wu}
\affiliation{%
  \institution{Huawei Technologies Co., Ltd.}
  \city{Shenzhen}
  \country{China}}
\email{wuxiaoyu7@huawei.com}

\author{Zheng Zheng}
\affiliation{%
  \institution{Beihang University}
  \city{Beijing}
  \country{China}}
\email{zhengz@buaa.edu.cn}
\authornote{Corresponding author.}

\begin{abstract}
Mobile crash repair requires more than identifying a source edit. Developers
must establish how the edit relates to the reported failure and whether the
modified application can be built, launched, and exercised without
disrupting valid behavior. In HarmonyOS applications, this assessment also
depends on platform-specific packaging and runtime execution.

To address this need, we present DEIA, an evidence-grounded workflow for
HarmonyOS crash repair. DEIA constructs a bounded source-evidence package,
uses a large language model to propose content-anchored edits, and relies
on a deterministic writer to materialize and check the patch. It derives
fail-to-pass (F2P) obligations from the reported fault and pass-to-pass
(P2P) obligations from the potential impact of the patch. A stage-level
ledger links the patch, test obligations, runtime observations, and
candidate-generation cost.

We evaluate DEIA on 30 repair tasks from 24 HarmonyOS applications and
compare it with an OpenHands-based workflow on 29 matched tasks. DEIA's
first-added-line location exactly matches the curated reference in 25/29
tasks (86.2\%), compared with 19/29 (65.5\%) for OpenHands, a difference of
20.7 percentage points. DEIA also uses 81.5\% fewer tokens, makes 60.8\%
fewer model calls, and takes 34.9\% less candidate-generation time on
average. Only two of the 58 behavior-specific P2P obligations were
executed, both for one repair pattern; comprehensive behavior preservation
therefore remains unestablished.
\end{abstract}



\maketitle

\section{Introduction}
\label{sec:intro}

Crash repair requires developers to interpret a failure report, locate the
responsible code, modify it, and assess the resulting behavior. Coding agents
such as OpenHands and SWE-agent can inspect repositories and edit source
files~\cite{wang2024openhands,yang2024sweagent}. Agentless organizes repair
around a more constrained sequence of steps~\cite{xia2024agentless}. Once a
patch has been generated, developers still need to trace it back to the crash
evidence that motivated the edit and forward to the package and replay used
to evaluate it. Each stage also incurs costs that must be measured separately.

In HarmonyOS applications written in ArkTS, a crash report typically
provides an exception and a stack trace.
However, it may not identify the complete source context needed for repair.
It may also lack a reliable reproduction sequence or a description of the
valid behaviors that should remain unchanged. Moreover, an ArkTS source modification must pass the platform
build and signing process to produce a HarmonyOS Ability Package (HAP). The
package must then be installed and exercised in a target runtime. These stages
produce different kinds of evidence. A patch may be applicable but still
fail to build. A package may build but fail to reach the reported trigger.
A replay may reach the trigger without showing that other valid behavior
has been preserved. Treating these outcomes as one repair-success signal
obscures what has actually been demonstrated. These requirements expose three connected challenges for an integrated
crash-repair workflow.

\paragraph{Challenge 1: Tracing Evidence Across Repair Stages.}
A failed replay can have several causes. The patch may still contain the
fault, the package may have failed to install, or the replay may never have
reached the trigger. Distinguishing these cases requires a record of which
source revision and patch produced the package, together with the outcome of
each stage from patch construction through replay.

\paragraph{Challenge 2: Grounding Behavioral Test Obligations.}
Removing the reported crash does not by itself show that the patch preserves
the application's valid behavior. A patch may suppress an exception while
also disabling a legitimate operation, changing a displayed value, or
altering a side effect. This concern is well recognized in automated program
repair, where a patch can satisfy available tests without implementing the
intended behavior~\cite{smith2015overfitting,ye2021patchassessment}. A repair
workflow therefore needs two complementary kinds of test obligations:
fail-to-pass (F2P) obligations that target the reported fault, and
pass-to-pass (P2P) obligations that target valid behavior potentially
affected by the patch. Each obligation needs an expected result supported by
an oracle, prerequisites for reaching the relevant action, and an observable
signal for judging the outcome. If the expected behavior or observation
method is unsupported, the obligation should remain pending rather than be
counted as a pass. Designed obligations must also be distinguished from
checks that have actually been executed.

\paragraph{Challenge 3: Measuring Stage-Specific Costs.}
A repair workflow consumes resources at several stages. These include model interaction, patch construction, package building, and device replay. Model calls, provider-reported tokens, and generation time characterize candidate generation. Build and replay time require separate measurements and instrumentation. Combining these quantities into one
undifferentiated cost can hide which stages were measured and which were not.
A meaningful comparison therefore needs a stage-level ledger that records
model-resource use within a defined boundary and keeps build and runtime
evidence separate when comparable timing is unavailable. Missing measurements
must remain unknown rather than being treated as zero.

These challenges motivate a common question.
\emph{How can a HarmonyOS repair workflow link each patch to the evidence
for its runtime behavior and account for the resources used to generate it?}
DEIA addresses the three challenges through three coordinated mechanisms.
The first mechanism establishes evidence provenance across repair stages.
The second grounds behavioral test obligations in supporting evidence.
The third records resource use within an explicit measurement boundary.

To address Challenge 1, DEIA separates source-context construction,
LLM-based diagnosis and edit planning, and deterministic patch
materialization. The evidence package is tied to an immutable source
revision. The model proposes content-anchored edits, while a deterministic
writer applies those edits to the original source and checks their anchors
and scope. The resulting patch is recorded with its provenance and carried
forward to package construction and device replay. This creates an auditable
link from the reported failure to the exact patch and package under
assessment. It also keeps patch applicability, build success, trigger
reachability, and runtime observations as distinct outcomes.

To address Challenge 2, DEIA derives F2P obligations from the reported
fault and P2P obligations from the potential impact of the proposed change.
Each obligation records an expected property, its supporting oracle
evidence, the prerequisites for reaching the relevant action, and the
observable signal used to judge the result. An obligation is eligible for
replay only when its expected property is supported and its outcome can be
observed. Otherwise, it remains pending. Eligible obligations receive a
fixed replay priority, beginning with the original trigger and then
considering high-impact valid behavior and adjacent or smoke checks. The
evaluation reports designed obligations separately from executed checks.
This distinction is important because an obligation specification does not
by itself demonstrate that the corresponding behavior was tested.

To address Challenge 3, DEIA records model calls, provider-reported input
and output tokens, and elapsed time through candidate-patch materialization
at $L_3$. Build and signing outcomes at $L_4$ and installation and replay
outcomes at $L_5$ are retained as separate evidence. Comparable matched
timing for these later stages is unavailable in the current study, so the
ledger does not treat missing measurements as zero or use them to claim a
reduction in end-to-end repair time.

The evaluation covers 30 repair tasks from 24 HarmonyOS applications.
One crash report describes a different defect from its reference patch,
leaving 29 matched tasks for comparison. Section~\ref{sec:dataset}
describes the collection, cohort construction, and public data release.
On those tasks, DEIA's first added line exactly matched a reference edit
location in 25 cases, compared with 19 for the OpenHands-based workflow.
DEIA also used fewer tokens and model calls and took less time to generate
a patch. The device records distinguish outcomes for the reported failure
from checks of other behaviors that the patch could affect.
Section~\ref{sec:rq2} reports which checks were executed. The three
contributions are summarized below.

\begin{itemize}[leftmargin=*,topsep=2pt,itemsep=3pt]
\item An evidence-grounded workflow connects crash reports, LLM-proposed
edits, deterministic patch construction, and device assessment. Its
stage-level records link each outcome to the patch under test. The
implementation and replication materials are available in the project
repository.\footnote{\url{https://gitcode.com/harmony-dfx-detection/answer_B/tree/DEIA}.}

\item A fault- and change-aware test-design protocol records the oracle,
prerequisites, and observation method for each F2P or P2P obligation. It
identifies checks with sufficient evidence for replay and records what is
missing for those that remain pending.

\item An empirical evaluation measures reference-location agreement,
patch applicability, device outcomes, and candidate-generation cost.
It reports the execution coverage of the designed behavioral checks.
\end{itemize}

\section{Problem Formulation}
\label{sec:problem}

\subsection{HarmonyOS Crash-Repair Task}

A HarmonyOS crash-repair workflow takes a faulty application revision and
a crash report as input, produces a candidate patch, and collects evidence
about patch applicability and application behavior. A repair task is represented as
\begin{equation}
    x = (S, \ell, d, \tau),
    \label{eq:repair-task}
\end{equation}
Here, $S$ denotes the immutable application source revision.
The variable $\ell$ represents the crash report, including the exception
and stack trace. The variable $d$ identifies the target runtime.
The optional sequence $\tau$ contains actions for reproducing the fault.
Given $x$, the repair workflow produces a candidate source
patch $P$.

Patch generation and repair validation are distinct objectives.
A candidate patch must first be applicable to the original
source revision. The modified application must then satisfy
the platform-specific build and signing requirements before
it can be installed and exercised in the target runtime.
These stages provide different levels of evidence. Patch
applicability establishes only that the modification can be
materialized, while successful building establishes that a
deployable package can be produced. Neither demonstrates
that the reported fault has been removed or that existing
behavior has been preserved.

Repair assessment therefore distinguishes patch applicability, buildability, and runtime behavior. Runtime evidence must also separate patched-side observations from paired fail-to-pass observations under comparable conditions. Similarly, a successful replay of one interaction
path does not establish the preservation of other valid
behaviors. The objective is consequently to maintain traceable
evidence across these stages rather than reduce repair success
to a single binary outcome.

\subsection{Behavioral Test-Obligation Formulation}

For a repair task $x$ and candidate patch $P$, two sets of behavioral
test obligations address complementary questions. The fail-to-pass (F2P)
set $\mathcal{O}^{F}$ targets the reported failure (Eq.~\eqref{eq:fault-obligations}).
The pass-to-pass (P2P) set $\mathcal{O}^{P}$ targets valid behavior the patch may affect
(Eq.~\eqref{eq:impact-obligations}). Their union $\mathcal{O}$ is the full set of
obligations for the patch (Eq.~\eqref{eq:all-obligations}).
\begin{subequations}\label{eq:obligation-generation}
\begin{align}
    \mathcal{O}^{F}
        &= \operatorname{FaultObligations}(\ell,S,\tau),
        \label{eq:fault-obligations}\\
    \mathcal{O}^{P}
        &= \operatorname{ImpactObligations}(S,P,K),
        \label{eq:impact-obligations}\\
    \mathcal{O}
        &= \mathcal{O}^{F} \cup \mathcal{O}^{P}.
        \label{eq:all-obligations}
\end{align}
\end{subequations}
The crash report $\ell$, source revision $S$, and any available reproduction
steps $\tau$ ground the F2P obligations in the reported trigger. For P2P
obligations, the patch identifies potentially affected behavior. The patch $P$ is interpreted against its original revision $S$. The
available behavior knowledge $K$ supplies evidence of its expected outcome from
application requirements, validated successful traces, or reviewed
specifications. Both sets use the same tuple for an individual obligation
\begin{equation}
    o = (u,\phi,\pi,\kappa,\omega),
    \label{eq:obligation}
\end{equation}
Here $u$ is the action to exercise and $\phi$ the property it should
satisfy. The oracle evidence $\pi$ supports that expectation independently
of the generated patch. The prerequisites $\kappa$ describe how to reach
the action, and $\omega$ specifies how to observe the property once reached.

A subscript $o$ selects a field of the current obligation. In
Eq.~\eqref{eq:eligible}, $\operatorname{Supported}$ tests whether
$\pi_o$ supports the property $\phi_o$.
The predicate $\operatorname{Observable}$ tests whether the method $\omega_o$
can detect that property. Equation~\eqref{eq:pending} collects the
obligations that fail either condition.
\begin{equation}
    E = \{o \in \mathcal{O} \mid
    \operatorname{Supported}(\phi_o,\pi_o) \land
    \operatorname{Observable}(\omega_o)\},
    \label{eq:eligible}
\end{equation}
\begin{equation}
    U = \mathcal{O} \setminus E,
    \label{eq:pending}
\end{equation}
The eligible set $E$ contains the obligations satisfying both predicates.
The pending set $U$ contains all remaining obligations. The design record
retains the missing evidence or observation method for each pending item.
Execution still requires reaching the prerequisites $\kappa$.

The test-design record $\rho_T$ pairs each obligation $o$ with its
design status $s_o$. The subscript $T$ identifies test design.
An item in $E$ has status \textsc{Eligible} and an item in $U$ has status
\textsc{Pending} (Eq.~\eqref{eq:design-record}).
\begin{equation}
    \rho_T =
    \{(o,s_o)\mid o\in\mathcal{O}\},
    \qquad
    s_o\in\{\textsc{Eligible},\textsc{Pending}\}.
    \label{eq:design-record}
\end{equation}
For eligible obligations, ordinal priorities P1--P3
determine their replay order. These priorities do not
introduce an adaptive test-selection policy or a separate
test-count budget. Importantly, obligation design and
eligibility are distinct from execution: an obligation
can only provide behavioral evidence after its
prerequisites are reached and its expected property
is actually assessed. Unreached triggers and unavailable
observations remain inconclusive rather than being
counted as successful checks.

\subsection{Cost Model}

Source preparation, model interaction, and device replay consume different
resources. To attribute a cost to a particular repair, the ledger links each
measurement to the source revision and the patch produced from it. The
generation interval ends when the candidate patch is materialized. Build
and device replay have separate records.

For each repair task $i$, the generation cost record contains three
quantities (Eq.~\eqref{eq:generation-cost}):
\begin{equation}
    C_i^{\mathrm{gen}} =
    \left(
        N_i^{\mathrm{calls}},
        N_i^{\mathrm{tokens}},
        T_{i,0:3}
    \right),
    \label{eq:generation-cost}
\end{equation}
The index $i$ identifies the task, and $\mathrm{gen}$ denotes candidate
generation. $N_i^{\mathrm{calls}}$ counts model calls made for that task.
The quantity $N_i^{\mathrm{tokens}}$ counts the total provider-reported input and output tokens.
The entry $T_{i,0:3}$ is elapsed time in seconds from the start of $L_0$
through the end of $L_3$. These stages cover evidence preparation, diagnosis,
edit proposal, and patch materialization, respectively. The later build/sign
and device-replay stages correspond to $T_{i,4}$ and $T_{i,5}$. These quantities
are maintained separately and are included in end-to-end
cost only when comparable measurements are available.
A missing measurement is treated as unknown rather than
zero. Consequently, $C_i^{\mathrm{gen}}$ characterizes
candidate-generation cost and does not represent the
complete cost of repair and runtime validation.

\section{Approach}
\label{sec:approach}\label{sec:overview}

\begin{figure}[!htbp]
\centering
\includegraphics[width=0.98\textwidth]{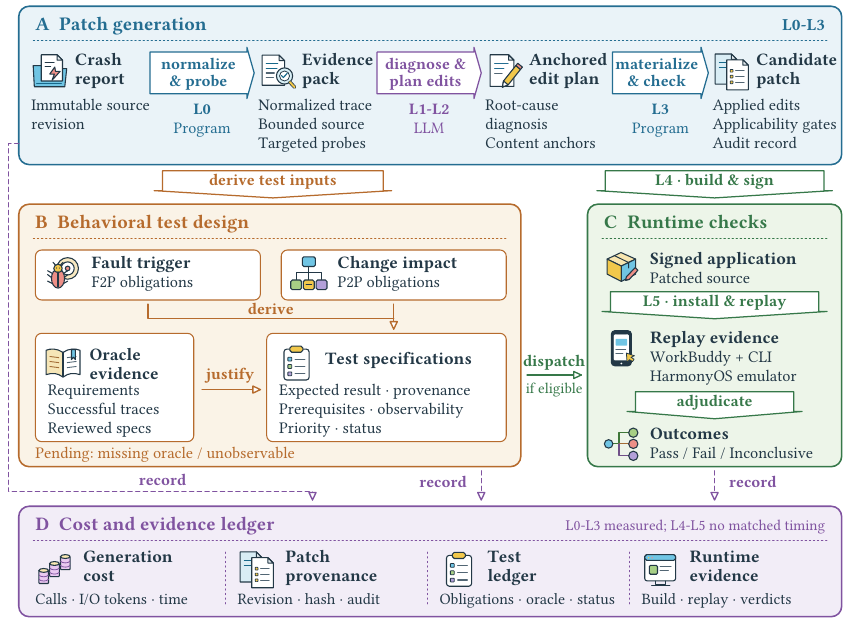}
\caption{Overview of DEIA. (A) Patch generation constructs a candidate patch from bounded source evidence and content-anchored edits. (B) Behavioral test design derives F2P/P2P obligations and specifies oracle-backed observations. (C) Runtime checks build, sign, install, and replay the candidate package. (D) The ledger links generation cost, patch provenance, test status, and runtime evidence. This study measures generation timing for $L_0$--$L_3$. Comparable matched timing for $L_4$--$L_5$ is unavailable.}
\Description{The overview contains four components. Patch generation (A) produces a candidate patch. Behavioral test design (B) derives F2P and P2P obligations and prepares test specifications. Runtime checks (C) build and sign the application, then install and replay it in a HarmonyOS emulator. The cost and evidence ledger (D) records generation cost, patch provenance, test status, and runtime evidence.}

\label{fig:overview}
\end{figure}

Figure~\ref{fig:overview} organizes DEIA into four components.
Patch generation (A) supplies the candidate patch and fault evidence.
Behavioral test design (B), expanded in Figure~\ref{fig:obligations},
specifies what to check. Runtime checks (C) build the candidate and replay
selected actions. The ledger (D) links the artifacts and measurements from
these components.

\subsection{Patch Generation}
\label{sec:patchgeneration}

\paragraph{$L_0$: Source Evidence.}
Figure~\ref{fig:overview}A begins with the crash report. DEIA parses
the report into stack frames and separates application
frames from framework frames. Source paths are resolved using progressively
longer suffix matches rather than only filenames. The resulting code set
is bounded. Optional static probes flag calls, identifier differences, and
possible dereferences. For cross-file producer--consumer faults, DEIA follows a deterministic
expression chain from the crash frame to related field references. It then
checks whether the bounded candidate set covers both the suspected producer
and consumer sides. This is a heuristic
retrieval and coverage gate, not a complete data-flow analysis. The evidence
package has a nominal 10{,}000-token context limit estimated from character counts.
Code near the crash frames has priority. The log records omitted files and
the difference between estimated and actual tokens. Better path resolution
increased the number of cases containing a reference-edited file from 25/30
to 29/30. This measures evidence coverage, not a proven increase in repair
success.

\paragraph{$L_1$--$L_2$: Diagnosis and Edit Proposal.}
The model proposes a cause and ranks possible edit locations. It returns
edits as \texttt{(file, old\_string, new\_string)} triples. DEIA records
the diagnosed location separately from the modified location. The model
decides what code to change. It does not write unified-diff headers or
line counts. Each \texttt{old\_string} must match exactly one location in
the original source. Section~\ref{sec:applicability} evaluates this edit
format for mechanical applicability, not semantic correctness.

\paragraph{$L_3$: Deterministic Materialization.}
A deterministic program applies the edits to the original source bytes
and constructs unified-diff patches. It checks that each edit anchor is
unique and each file path is allowed. The patch files are written as bytes
to preserve their line endings. The current generated patches pass
\texttt{git apply --check} in all 30 tasks. The delivered patch artifacts
are byte-identical to the generated patches, as verified by SHA-256, and
all 30 delivered artifacts pass the same check.

\subsection{Behavioral Test Design}
\label{sec:testdesign}
\begin{figure}[!htbp]
\centering
\includegraphics[width=0.94\textwidth]{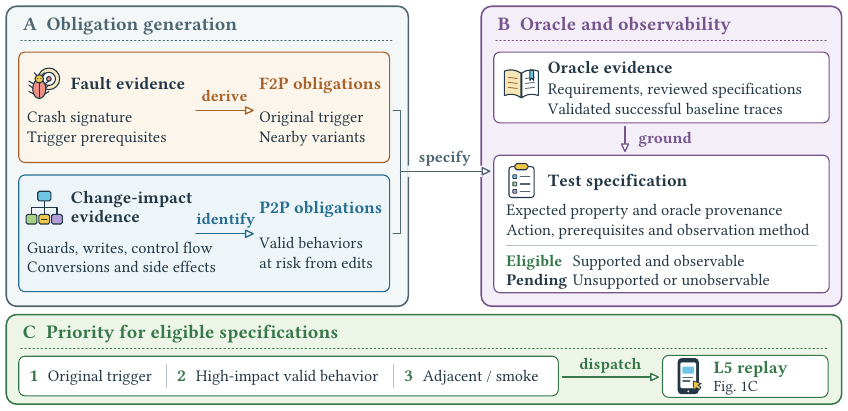}
\caption{Detail of the behavioral test-design component in Figure~\ref{fig:overview}B. (A) Fault and change-impact evidence derive F2P and P2P obligations. (B) Oracle evidence grounds each expected property, while the specification records the action, prerequisites, and observation method. Supported and observable specifications are eligible. Unsupported or unobservable specifications remain pending. (C) Eligible specifications receive an ordinal replay priority and are dispatched to $L_5$ in Figure~\ref{fig:overview}C.}
\Description{Panel A derives F2P obligations from fault evidence and P2P obligations from change-impact evidence. Panel B connects oracle evidence to a test specification containing an expected property, prerequisites, and an observation method. Supported and observable specifications are eligible. Unsupported or unobservable specifications remain pending. Panel C assigns replay priority to eligible specifications and dispatches them to L5 replay in Figure 1C.}

\label{fig:obligations}
\end{figure}

Figure~\ref{fig:obligations} provides the operational detail for the
behavioral test-design component in Figure~\ref{fig:overview}B. The
following paragraphs describe its three steps in order. Test authors may
supply or review expected properties and prerequisites.

\paragraph{A. Derive fault and change-impact obligations.}
\label{sec:f2pdesign}

F2P checks start from the reported crash. The design records its exception,
likely trigger, and required device or UI state. It may also cover nearby
cases, such as invalid indices or missing values, when the fault evidence
supports them. The report states whether the original faulty version was
replayed under comparable conditions. A historical crash log is not such
a replay. If the device cannot reach the trigger, the result is inconclusive.

\label{sec:p2pdesign}

P2P checks focus on valid behavior that a patch might change. A new
guard or return statement may affect normal execution. A default value or
data conversion can also change valid behavior. The designer considers how an overly broad fix could
remove a crash but break a valid path. Each expected result must be supported
independently of the patch. A requirement or a verified earlier trace can
provide that support. For example, an index-guard modification can motivate an invalid-index F2P
replay and a valid reorder or selection P2P check. The expected behavior
must be grounded in the valid-input and list-order contract, not inferred
from the patch itself. This is a design example, not evidence that either
check was executed.

\paragraph{B. Specify the oracle and observation method.}
\label{sec:allocation}

The protocol records five fields for each obligation: its action,
prerequisites, expected property, oracle provenance, and observation method. An obligation is
eligible only when its expected property has supporting evidence and its
effect can be observed. Otherwise, it remains pending, with the missing
support or observation requirement retained in the design record. Reaching a screen establishes reachability. It does not establish that a
state update, value, or side effect was preserved.

\paragraph{C. Assign replay priority.}
Among eligible specifications, the original crash trigger is the mandatory
P1 design target and receives first replay priority. P2 covers high-impact
valid behaviors that may be affected by the patch. P3 covers adjacent paths
and smoke checks. These labels express an ordinal replay order, not a learned
risk score, a test-count budget, or an adaptive allocation algorithm. The
released suite uses a fixed eight assertions per task (two F2P and six P2P),
for 240 design slots across 30 tasks. The ordinal labels guide replay planning
without changing that fixed suite. 

Algorithm~\ref{alg:testdesign} retains the missing evidence for each pending decision. Some expected properties in this study drew on curated reference repairs
and human judgment. The specification records that provenance so that a
reviewer can assess the independence of the oracle.

\begin{algorithm}[t]
\caption{Procedure for deriving and recording F2P/P2P test obligations.}
\label{alg:testdesign}
\begin{algorithmic}[1]
\Require task $x=(S,\ell,d,\tau)$, candidate patch $P$, behavior knowledge $K$
\State initialize $E$, $U$, and $\rho_T$ as empty sets
\State $\mathcal{O}^{F}\gets\operatorname{FaultObligations}(\ell,S,\tau)$
\State $\mathcal{O}^{P}\gets\operatorname{ImpactObligations}(S,P,K)$
\ForAll{$o\in \mathcal{O}^{F}\cup\mathcal{O}^{P}$}
  \State record obligation type, action $u_o$, prerequisites $\kappa_o$, and observation method $\omega_o$
  \State record expected property and supporting oracle source, if available
  \If{$\operatorname{Supported}(\phi_o,\pi_o)\land\operatorname{Observable}(\omega_o)$}
    \State set $s_o\gets\textsc{Eligible}$
    \State assign ordinal replay priority and add $o$ to $E$
  \Else
    \State set $s_o\gets\textsc{Pending}$
    \State record missing evidence and add $o$ to $U$
  \EndIf
  \State add $(o,s_o)$ to $\rho_T$
\EndFor
\State \Return $(E,U,\rho_T)$
\end{algorithmic}
\end{algorithm}

\subsection{Runtime Checks}
\label{sec:runtime}

The runtime checks in Figure~\ref{fig:overview}C take the patch from A and
the selected checks from B. At $L_4$, the patch is applied to a clean source tree. HarmonyOS
tools then build and sign a HAP. Build and dependency failures are recorded
separately. At $L_5$, the replay harness uses the HarmonyOS command-line and
emulator interfaces to install and launch the package. It performs UI
actions, replays triggers, and collects logs and process/UI signals. For
each check, the record identifies the requested action and whether it
was reached. It then records the observed result. If execution is blocked,
the record states why.

\paragraph{Interpreting the observations.}
\label{sec:evidencelevels}
The result record links each observation to its stage. A build establishes
that the toolchain can produce a package. A paired F2P result requires a
failure on the original version and a successful replay under comparable
conditions on the patched version. A P2P observation applies to the input
and property that the check actually reached. Unreached triggers remain
inconclusive, and infrastructure failures retain their stage-specific cause.
Section~\ref{sec:metrics} defines the measures and Section~\ref{sec:rq2}
reports their coverage.

The feedback policy identifies nine failure types. These include missing
or ambiguous edit anchors and edits outside the approved scope. Other
categories cover patch failures and compilation problems. The policy also
records signing or installation rejection, unreached triggers, and
infrastructure problems. The model can use feedback about fixable edits
and compilation errors. Setup failures or unobservable device behavior
should not be treated as errors in the patch itself. This policy describes
control flow. The study does not measure a retry improvement from it.

\subsection{Cost and Evidence Ledger}
\label{sec:methodcost}
Figure~\ref{fig:overview}D organizes the ledger into four linked record
groups. They form a provenance chain from candidate generation to device
assessment. The ledger is not a single success flag. It preserves the
evidence and status associated with each stage.

For each repair, the patch record ties the source revision to the model's
edit anchors and the hash of the patch actually built. The associated cost
record measures model calls, input and output tokens, and elapsed time up
to patch materialization at $L_3$. Later build and replay records retain
their outcomes. Comparable timing for those stages was unavailable.

The test ledger links each expected property to its evidence, prerequisites,
and observation method. It distinguishes checks ready for replay from
pending checks and records any human contribution to their specification.
The runtime record then identifies the tested package, the action reached,
and the observed outcome. Together these records let a reader distinguish
a failure to install the package from a recurrence of the target crash or
a change in the behavior being checked.

\section{Experimental Design}
\label{sec:experimentaldesign}

\subsection{Dataset Construction}
\label{sec:dataset}

The tasks come from a HarmonyOS crash-repair collection released for a
software-defect challenge. The collection and its metadata are publicly
available at \url{https://gitcode.com/dfx-contest-2026/data}. The collection
contains 70 reported crashes. Grouping the reports by crash signature and top application
stack frame produced 33 candidate fault instances. We excluded three
instances because their reported defects were absent from the supplied
source revisions. This left a \emph{generation cohort} of 30 tasks from
24 applications.

One task has a crash report that does not describe the defect addressed by
its curated reference repair. This task contributes to the 30-task generation and
cost statistics. The primary localization and device comparisons exclude it. The resulting \emph{adjudicable cohort} contains 29 tasks.
We identified this mismatch during analysis and made the exclusion
decision post hoc. Reporting both cohorts makes the
post-hoc exclusion visible and keeps generation statistics separate from
matched comparisons. Table~\ref{tab:dataset} describes the 30 tasks retained for generation. Device-execution accommodations are described in
Section~\ref{sec:settings}.

\begin{table}[!htbp]
\centering
\begin{minipage}{0.82\linewidth}
\centering
\caption{Composition of the 30 repair tasks from 24 HarmonyOS applications.}
\label{tab:dataset}
\begin{tabular*}{\linewidth}{@{\extracolsep{\fill}}llrr@{}}
\toprule
\textbf{Characteristic} & \textbf{Category} & \textbf{Tasks} & \textbf{Share (\%)} \\
\midrule
Data batch & B1 & 12 & 40.0 \\
           & B2 & 8 & 26.7 \\
           & B3 & 10 & 33.3 \\
\addlinespace[2.4pt]
Exception type & TypeError & 16 & 53.3 \\
               & RangeError & 9 & 30.0 \\
               & SyntaxError & 5 & 16.7 \\
\addlinespace[2.4pt]
Log source & Device-origin & 24 & 80.0 \\
           & Augmented & 6 & 20.0 \\
\addlinespace[2.4pt]
Reference patch & Single-file & 27 & 90.0 \\
                & Multi-file & 3 & 10.0 \\
\bottomrule
\end{tabular*}
\end{minipage}
\end{table}

\paragraph{Data leakage and workspace isolation.}
\label{sec:leakage}

The model-visible crash reports were stripped of embedded text that
could disclose root-cause explanations. For DEIA, the intended model input
was the generated evidence package and permitted source excerpts. Reference
fixes, task answers, and test scripts were not intended to be model-visible.
The OpenHands baseline configuration excluded the \texttt{testcases/} and
\texttt{reports/} directories.

Strict workspace isolation was not enabled for the full DEIA batch:
the \texttt{--isolated} mode was validated on one case but not used for all
30 tasks. A post-run audit covered 175 model sessions containing about 25.7 million
characters. The audit found no matches for the recorded answer-leak markers. This
audit reduces, but does not eliminate, the risk of unintended exposure.
Earlier development also involved prompt changes on overlapping cases.
The 30 cases are therefore a study of the developed workflow, not a held-out
estimate of generalization. A future evaluation needs untouched cases.

\subsection{Research Questions}
\label{sec:researchquestions}

\begin{itemize}[leftmargin=*,topsep=2pt,itemsep=3pt]
\item \textbf{RQ1. Candidate generation.} How often does each workflow edit
  a reference location and produce a patch that applies to the original source?
\item \textbf{RQ2. Device validation.} What do build and replay records show
  about removal of the reported failure and preservation of affected behavior?
\item \textbf{RQ3. Generation cost.} How do the workflows compare in model
  calls, token use, and elapsed time through patch materialization?
\end{itemize}

\subsection{Baselines}
\label{sec:baselines}

The primary external baseline is an OpenHands-based repair workflow. The team's earlier
plugin-based workflow provides an additional workflow-level reference. These
comparisons do not isolate individual components unless a metric is explicitly
tied to a common procedure.

\paragraph{OpenHands-Based Autonomous Repair.}
The OpenHands baseline follows a sandboxed autonomous-agent loop based on
the OpenHands framework~\cite{wang2024openhands}. Given the crash report and
application source tree, the agent explores the repository and edits source
files through Bash. Its available actions are Bash and task completion.
After the agent finishes, \texttt{git diff} extracts the resulting patch
from the sandbox working tree. The current comparison uses the same 30-task generation
cohort as DEIA: the 30 eligible tasks remaining after excluding the three
reports whose alleged defects were absent from the supplied source revisions.
The crash-report bytes and source revisions are matched at the task level. Reference patches, test scripts, and evaluation
reports are excluded from the agent workspace.

The baseline implementation was maintained in a separate
\texttt{dfx-patch-agent/} workspace. The available design record describes
12 modules, but the workspace was not version-controlled and is not included
in the public DEIA release. The released materials do not pin an exact upstream OpenHands version or
commit. The term \emph{OpenHands-based} identifies the implementation
evaluated here. Archived per-task records and patches provide the evidence
for this comparison. The public artifact does not currently support an
independent rerun of this baseline.

\paragraph{Previous Plugin-Based Repair Workflow.}
The second comparator is the team's earlier plugin-based repair pipeline.
It asks the model to produce a diagnosis and a unified-diff patch, which is
then checked by surrounding scripts. It shares evidence construction, the fault-mode knowledge base, and some
function-level replay utilities with DEIA. However, the workflows differ in
four ways. The model writes a unified diff directly instead of returning
structured edit operations. The tool surface includes Bash. The interaction
and retry strategy uses multiple rounds. Stack-frame paths are handled
differently from DEIA's path-suffix matching. Because several factors change
together, this is a workflow-level comparison rather than an isolated
ablation of patch representation, evidence construction, or tool restriction.

\paragraph{Role of Curated Reference Patches.}
The reference patches were manually prepared during dataset curation and
refined as the cases became better understood. They are not necessarily
fixes committed by upstream developers. Both workflows excluded the reference patches from their declared
generation inputs. They were also excluded from DEIA's bounded evidence
package. Workspace isolation and the limits of the post-run exposure audit
are reported in Section~\ref{sec:leakage}. The evaluation uses these patches
to score edit-location agreement offline. Some behavioral expectations also
drew on reference repairs, as discussed in the construct-validity analysis. A short upstream change may differ substantially from a more
defensive reference patch. Reference-line coverage must therefore not be
treated as correctness.

Device validation uses packages built from each method's own generated
patch: DEIA packages come from \texttt{patches/deia/}, and OpenHands
packages come from the OpenHands working-tree diffs. No gold-reference
package is counted as a DEIA result. The curated reference patch is not
treated as a fully independent behavioral oracle. Its provenance and
possible defects are discussed in the validity analysis.

\subsection{Experimental Settings}
\label{sec:settings}

Both workflows ran GPT-6.1-Sol through DeepSeek Harness (DSH). For a
given task they received the same crash report and source revision.
DEIA supplied bounded source evidence and accepted structured edits.
The OpenHands-based workflow used Bash to explore and modify its sandbox.
Token counts come from the usage reported by DSH.

Package construction and validation use the HarmonyOS command-line
toolchain. Its components include \texttt{ohpm}, \texttt{hvigorw}, signing
utilities, and \texttt{hdc}. Device checks use an API-26 HarmonyOS x86\_64
emulator with a 12-second observation window. Both comparison arms use the
same replay harness and platform interfaces. On November 6, 2026, packages
from both workflows ran sequentially on the same emulator instance. They
used a byte-identical harness and the same scenario definitions. Most DEIA packages had been built
before the matched replay. One package required an unavailable platform
dependency and was rebuilt with the same local compatibility stub in both
arms. Six tasks required native libraries. Both arms received the same
x86\_64 library injection and re-signing adjustment. These adjustments address
emulator compatibility and do not modify the source lines changed by either
repair patch.

\subsection{Evaluation Metrics}
\label{sec:metrics}

The generation cohort supplies 30 tasks for artifact checks. The
matched comparison uses 29 tasks to assess edit locations, device
observations, behavioral coverage, and generation cost. The following
metrics specify the evidence used for each outcome.

\subsubsection{Source-Level Localization}

File-level hit measures whether a generated patch modifies a file also
modified by the curated reference patch. The first-added-line metric
compares the first added line in the generated patch with added-line
locations in the reference patch, using the new-file line number. The comparison measures
exact agreement and agreement within two and five lines. These are
reference-alignment measures, not semantic correctness metrics. An
additional any-changed-line analysis is reported separately where relevant.

\subsubsection{Patch Applicability}

Patch applicability is determined by running strict
\texttt{git apply --check} against the original source revision. For DEIA,
the check is applied to the current generated patches and repeated on the
delivered patch artifacts. SHA-256 comparison verifies byte identity
between the generated and delivered files for all 30 tasks. Applicability
establishes whether a patch can be applied. It does not establish that the
resulting program is correct.

\subsubsection{Device and Test-Obligation Coverage}

Each task has two F2P assertions and six P2P assertions. This gives 240
assertions for the 30-task generation cohort and 232 for the 29 matched
tasks. Each matched task contributes one F2P check on the original trigger
and one on a nearby trigger. Its six P2P checks cover a precondition, two
specific behaviors, an adjacent path, and two smoke checks.
Figure~\ref{fig:coverage} gives the denominator for each category.

Device adjudication, runtime fault confirmation, paired F2P outcomes, and
reached P2P-derived checks are reported separately. A compile-gate outcome
is not counted as a reproduced runtime crash. A patched-side observation
is not treated as complete pass-to-pass evidence unless the corresponding
unpatched behavior and expected property are established. Designed,
executed, and adjudicated counts therefore retain separate denominators.
Coverage counts are not behavioral pass rates.

\subsubsection{Candidate-Generation Cost}

Generation cost comprises model-call counts, token use, and elapsed time
through candidate patch materialization at $L_3$. Token use is recorded as
input, output, and total tokens. The cost boundary excludes package building,
signing, installation, and device replay. The DEIA-only stage-time decomposition
uses the 30-task generation cohort. The DEIA--OpenHands cost comparison
uses the 29-task matched cohort. Values are point estimates from the
recorded runs. No repeated-run variance is available.

\subsection{Execution Protocol}
\label{sec:execution}

Each workflow generated one candidate per task from the same crash report
and source revision in its own workspace. For static evaluation, we
compared each candidate's edit locations with the curated reference
and checked its applicability against the original source. For DEIA,
the check covered all 30 current generated patches and their corresponding
delivered artifacts. SHA-256 comparison verified that each delivered
artifact was byte-identical to its generated patch. The same applicability
check was applied to the OpenHands working-tree diff on the matched tasks.

For the 29 matched tasks, the replay harness tested packages built from
each workflow's own patch. Both sets ran sequentially on the same API-26
emulator using \texttt{repro\_all\_cases.py} and the same 12-second
observation window. The harness recorded the installed package, the action
reached, and the observed fault or behavior. Pairing these records by task
allowed the comparison to use a common baseline for each reported failure.

\paragraph{Reproduction Scope.}
The DEIA GitCode branch provides the implementation and prompts alongside
the task metadata, patches, offline metrics, and replay records. The
original application archives are not included in the paper package. Reproduction
requires the archived source revision and the recorded runtime configuration,
including DSH \texttt{0.1.5-rc.2}, the HarmonyOS toolchain, and the emulator.
The OpenHands-based results rely on archived outputs because its workspace
was not released and its upstream commit was not pinned.

\FloatBarrier
\section{Results}
\label{sec:results}

\subsection{RQ1: Candidate-Generation Reliability}
\label{sec:rq1}

RQ1 examines reference-aligned localization and mechanical patch
applicability. These measures describe where a workflow edits and whether
its patch can be applied to the original source. Neither measure alone
establishes that the resulting application behaves correctly.

\subsubsection{Reference-Aligned Localization}
\label{sec:localization}

\begin{figure}[!htbp]
\centering
\includegraphics[width=\linewidth]{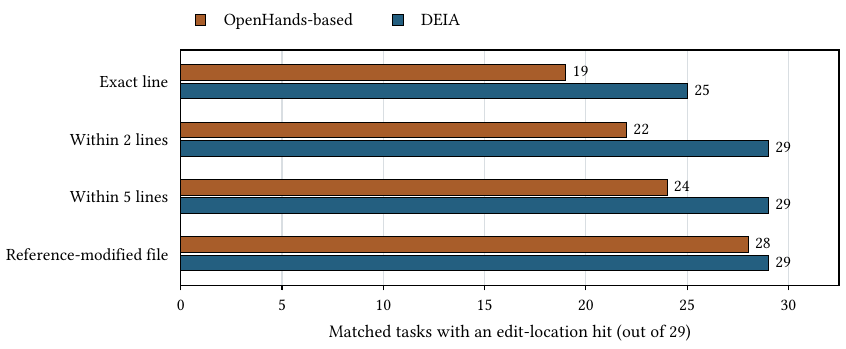}
\caption{Agreement with reference edit locations on 29 matched tasks. Each pair of bars compares OpenHands-based and DEIA counts. The first three rows assess the first added line at increasing tolerances. The last row counts edits to any file changed by the reference patch.}
\label{fig:localization}
\Description{Grouped horizontal bars compare OpenHands-based and DEIA. The counts are 19 and 25 for exact first-added-line agreement, 22 and 29 within two lines, 24 and 29 within five lines, and 28 and 29 for reference-modified files.}
\end{figure}

Figure~\ref{fig:localization} compares the two workflows on the 29 matched
tasks. For line-level agreement, the first added line of each generated patch
is compared with the added lines in the curated reference using new-file
line numbers.

OpenHands modified a reference-edited file in 28 of 29 tasks. DEIA did so in all 29. Exact first-added-line matches occurred in 19 OpenHands patches and 25 DEIA patches. This corresponds to 65.5\% and 86.2\%, respectively, or a 31.6\% relative increase for DEIA. At a two-line tolerance, DEIA matches 29 tasks and OpenHands matches 22.

The line-level metric is sensitive to the selected patch anchor. Under an
\emph{any changed line} definition, exact overlap is 23/29 for OpenHands
and 26/29 for DEIA. Within two lines, the results are 25/29 and 29/29.
The main comparison retains the first-added-line metric specified in the evaluation protocol. Several misses arise
from different edit boundaries within the same code block rather than a
different function. The metric therefore captures agreement about where to edit.
Assessing the diagnosis itself would require evidence about the fault's cause.

The DEIA evidence package contains a reference-edited file for 29/30 tasks
after path-suffix matching, compared with 25/30 before that correction.
This is evidence-package coverage and is separate from generated-patch
localization.

\subsubsection{Cross-File Identifier Mismatch}
\label{sec:crossfilecase}

Two reports in the cohort illustrate why a crash frame does not always
identify the source location that should be edited. Both reports point to an undefined
\texttt{authorNickName} dereference in the
\texttt{ArticleFeedDetails.ets} file, at line 111. The underlying cause is upstream:
\texttt{MockArticle.ets} assigns \texttt{Author\_2}, whereas the available
mock user is keyed by \texttt{author\_2}. A strict equality lookup in \texttt{AuthorServiceMock.ets} therefore
returns no author. The data service leaves the author field unset. The view
later dereferences that field.

For these cases, DEIA starts from \texttt{this} and follows
\texttt{newsFeedVM} to \texttt{articleDetails}. It then traces the
\texttt{author} field to \texttt{authorNickName}. The three-hop search
checks producer/consumer coverage.
The resulting 15-file focus set places the crash file first and
\texttt{MockArticle.ets} second. The generated patch changes the
producer-side identifier at \texttt{MockArticle.ets:50} from
\texttt{'Author\_2'} to \texttt{'author\_2'}. It does not suppress the
view-side exception with a null fallback. The replay reaches the trigger
and removes the target crash. The device showed the expected author name, and the replay recorded no new fault. An earlier view-only fallback patch either left
the target failure or hid the author value. This illustrates the difference
between removing a symptom and repairing the producer--consumer mismatch. The two reports share the same patch bytes and each has one recorded
repeat. The two observations therefore cover one repair pattern.

\subsubsection{Mechanical Patch Applicability}
\label{sec:applicability}

Table~\ref{tab:rq1-summary} reports strict
\texttt{git apply --check} results for the generated and delivered
artifacts.

\begin{table}[!htbp]
\centering
\begin{minipage}{0.86\linewidth}
\centering
\caption{Mechanical patch applicability by workflow and artifact.}
\label{tab:rq1-summary}
\begin{tabular*}{\linewidth}{@{\extracolsep{\fill}}llrrr@{}}
\toprule
\textbf{Workflow} & \textbf{Patch artifact} & \textbf{Tested} & \textbf{Passed} & \textbf{Rate (\%)} \\
\midrule
Plugin-based & Generated diff & 30 & 14 & 46.7 \\
DEIA & Generated and delivered patch & 30 & 30 & 100.0 \\
OpenHands-based & Working-tree diff & 29 & 29 & 100.0 \\
\bottomrule
\end{tabular*}
\deiatablenote{Passed counts patches accepted by \texttt{git apply --check}
against the corresponding original HEAD. The DEIA row covers all 30
generated tasks, including the 29-task matched subset. DEIA's generated
and delivered patches are byte-identical in all 30 tasks (SHA-256).
On the matched subset, both DEIA and OpenHands pass 29/29.
Applicability does not establish buildability or behavioral correctness.}
\end{minipage}
\end{table}

DEIA generates diff headers and line counts programmatically rather than
asking the model to write them. The current generated patches pass strict
applicability checks in all 30 tasks. The delivered patch artifacts are
byte-identical to the generated patches in all 30 tasks, as verified by
SHA-256, and all 30 delivered artifacts also pass the check. On the same
29-task adjudicable subset, both DEIA and OpenHands pass 29/29. The results
therefore support reliable patch materialization and delivery, not an
applicability advantage over OpenHands.

The previous plugin-based workflow differs from DEIA in tool access,
source-path handling, and retry policy as well as patch representation.
Its 14/30 result is therefore a workflow-level comparison, not an isolated
ablation of structured edits.

\begin{rqanswer}{Answer to RQ1.}
On the 29 adjudicable cases, DEIA has higher reference-aligned file and
first-added-line scores than the OpenHands-based workflow. Both methods'
patches pass strict applicability checks on this matched subset. All 30
current DEIA generated patches and their byte-identical delivered artifacts
also pass. These results support a difference in reference-location
agreement, not a claim of superior semantic correctness or greater patch
applicability than OpenHands.
\end{rqanswer}

\begin{implicationbox}
\paragraph{Implication.}
Patch applicability, source-location agreement, and repair behavior are
separate properties. Deterministic materialization and byte-verified
delivery make the current patch artifacts reproducibly applicable.
Reference alignment and applicability alone do not establish semantic
correctness.
\end{implicationbox}

\FloatBarrier
\subsection{RQ2: Device Validation Outcomes}
\label{sec:rq2}

The matched device comparison uses packages generated by the two methods:
DEIA packages are built from its method-generated patch set, and OpenHands
packages are extracted from the OpenHands-based agent's working trees.
Curated reference patches supply the offline localization targets. Each
device outcome is attributed to the workflow that generated the tested patch.

\subsubsection{Device Outcomes}

On November 6, 2026, both workflows' patched packages ran on the same
API-26 emulator with a 12-second observation window. The DEIA manifest
links every tested package to its generated patch. All 29 generated patches
differ from their curated references. In both arms the replay reached the
patched target path in every task. Processes survived the observation
window, cleanup succeeded, and the harness recorded no new faults or
navigation errors.

Interpreting these outcomes requires the baseline for each task. The
unpatched application reproduced a runtime fault in 28 cases. In the
remaining case the original version failed to build, so its outcome records
a transition from build failure to build success. One of the 28 runtime
cases required a local compatibility stub on the emulator. Both method
packages used the same stub, which left the edited source line untouched.
The other 27 cases supplied patched-side runtime observations.

The target-path judge paired each method's patched outcome with the
baseline for the same source revision and trigger. This yielded 29 target
adjudications per workflow, including the build-gate case. The alternative
trigger was never replayed for any of the 29 tasks, so its outcome remains
unknown.

\subsubsection{P2P-Derived Checks}

Each method's patched packages produced 87 observations across three
phases: reaching the precondition, cold-start rendering, and return
navigation. The two methods had the same outcomes on these checks.
The original behavior was not established for every corresponding input,
so these observations support reachability and smoke behavior on the
patched paths.

\subsubsection{Obligation Coverage}

\begin{figure}[!htbp]
\centering
\includegraphics[width=0.98\textwidth]{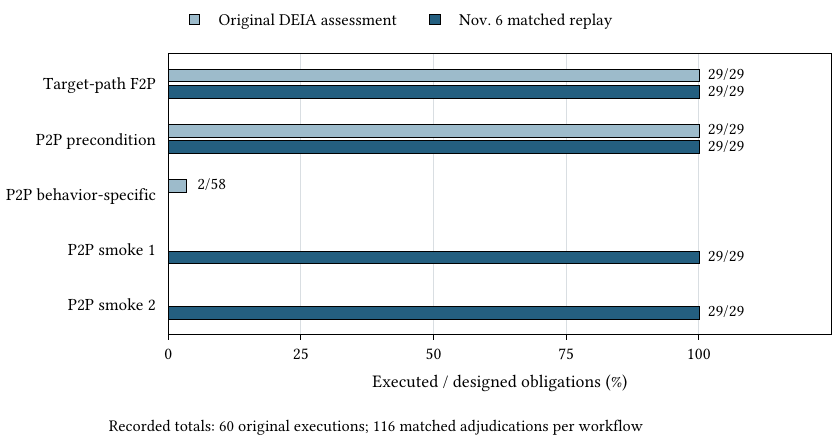}
\caption{Recorded executions of selected F2P and P2P-derived obligations on the 29-task adjudicable cohort. The figure includes only phases with recorded executions. The original DEIA assessment contains 60 records. The November 6 matched replay contains 116 adjudications per workflow.}
\Description{The plot shows executed target-path F2P and P2P precondition checks in both assessments. It also shows two behavior-specific P2P executions in the original DEIA assessment and two P2P smoke phases in the matched replay. The original assessment contains 60 records, and the matched replay contains 116 adjudications per workflow.}
\label{fig:coverage}
\end{figure}

The matched cohort contains 232 designed assertions across 29 tasks.
Figure~\ref{fig:coverage} compares the phases with recorded checks in
the original DEIA assessment or the November 6 replay. The original assessment recorded
60 executions. The matched replay recorded 116 adjudications per workflow.
The matched replay did not separately run the behavior-specific checks.
The two smoke phases were outside the original assessment.

Across the 58 designed behavior-specific P2P obligations, only two have
been executed. Both observations come from reports that share the same
producer--consumer identifier-mismatch repair pattern. They use equivalent
UI-text observations. They do not establish a general functional-correctness
rate or constitute independent replications. All 29 adjacent-path P2P
obligations and all 29 near-path F2P obligations remain unexecuted.

The judge was also applied to pre-ABI-fix installation-failure records.
It returned 75 PASS, 8 FAIL, and 7 INCONCLUSIVE. The judge can therefore
distinguish some blocked setup records. The failures in this control arise
from installation and navigation blockage, not controlled semantic
regressions. This validates some infrastructure sensitivity. It does not
establish broad behavioral-regression sensitivity.

A controlled negative-patch dry run illustrates the distinction between
F2P and P2P. The negative patch widens an index guard by one position.
The target crash remains absent. A valid drag to the final position should
change the list order, but the recorded order stays the same.
The corresponding P2P check therefore fails while F2P remains pass. This
single dry-run example is not a broad sensitivity estimate. Its driver has
not completed device execution.

\begin{rqanswer}{Answer to RQ2.}
Both methods receive 29/29 device-adjudication outcomes on their own
packages under the same-day API-26 replay. Each arm also receives 29/29
target-obligation adjudications and 87/87 P2P-derived reachability and smoke
checks. The 29 adjudications include 28 runtime fault cases and one
compile-gate case. One task uses the same compatibility stub in both arms.
The 87 checks are patched-side observations, not complete pass-to-pass tests.
The original DEIA assessment executed two of the 58 behavior-specific P2P
obligations. Near-path F2P and
adjacent-path P2P remain unexecuted. The evidence supports comparable outcomes
on checked paths, not comprehensive functional correctness, formal
equivalence, or universal repair superiority.
\end{rqanswer}

\begin{implicationbox}
\paragraph{Implication.}
Device validation should report package provenance and distinguish trigger
reachability from preservation of other valid behavior. Patched-side
reachability and smoke observations provide evidence about selected
execution paths, but they do not replace independently oracled P2P
regression tests.
\end{implicationbox}

\FloatBarrier
\subsection{RQ3: Candidate-Generation Cost}
\label{sec:rq3}

\begin{figure}[!htbp]
\centering
\includegraphics[width=\linewidth]{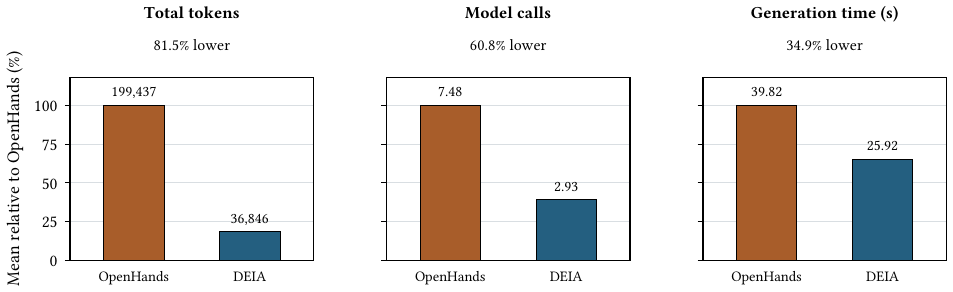}
\caption{Candidate-generation cost on 29 matched tasks. Each panel normalizes the OpenHands-based mean to 100. Labels above the bars show absolute means, while the percentage above each panel reports DEIA's reduction relative to OpenHands. Each workflow ran once per task.}
\label{fig:generation-comparison}
\Description{Three panels compare total tokens, model calls, and generation time. In each panel, the OpenHands-based mean is normalized to 100. DEIA has normalized values of 18.5, 39.2, and 65.1, respectively. Absolute means are printed above the bars. The reductions are 81.5 percent, 60.8 percent, and 34.9 percent.}
\end{figure}

On the 29 matched tasks, Figure~\ref{fig:generation-comparison} compares
resource use for generating a candidate patch. The separate timing
breakdown uses all 30 DEIA tasks to identify where generation time was
spent. Both analyses end at patch materialization, before build and replay.

Most of the token difference comes from input: DEIA averaged 36,183 input
tokens per task, compared with 198,819 for OpenHands (81.8\% fewer).
Mean output tokens rose from 618 to 663 (7.3\% more). Median generation
time fell from 35.15 to 22.93 seconds (34.8\%). These workflow-level
measurements do not isolate the effect of any one design choice.

The previous plugin-based workflow is a secondary workflow-level
reference. It uses 101,882 tokens, 7.10 model calls, and 102.47 seconds
per task in its recorded 30-task run. Its tool access, path handling, and
retry policy differ from DEIA, so these values are descriptive context
rather than a controlled ablation.

\paragraph{Generation-time breakdown.}

\begin{figure}[!htbp]
\centering
\includegraphics[width=0.94\textwidth]{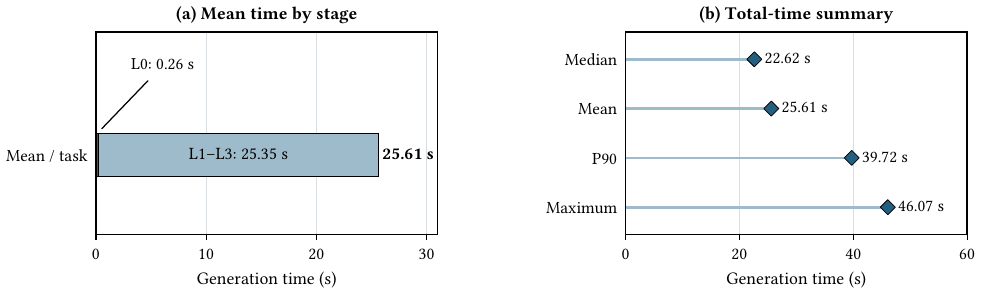}
\caption{Candidate-generation time for DEIA on the 30-task generation
cohort. (a) Mean time for local evidence construction ($L_0$) and model
interaction, editing, and checks ($L_1$--$L_3$). (b) Median, mean, P90, and
maximum generation time. Both panels exclude package building and device
replay.}
\Description{Panel A shows a mean generation time of 25.61 seconds.
Local evidence construction (L0) accounts for 0.26 seconds. Model interaction,
editing, and checks (L1--L3) account for 25.35 seconds. Panel B reports a
median of 22.62 seconds, a mean of 25.61 seconds, a P90 of 39.72 seconds,
and a maximum of 46.07 seconds.}
\label{fig:cost}
\end{figure}

Figure~\ref{fig:cost} summarizes DEIA's 30-task generation-time record.
The mean is 25.61 seconds: 0.26 seconds for local evidence construction
($L_0$) and 25.35 seconds for model interaction, edit materialization, and
checks ($L_1$--$L_3$). This figure uses the 30-task DEIA generation cohort.
The 29-task comparison in Figure~\ref{fig:generation-comparison} excludes the mismatched
crash-report/reference-repair task. The stage breakdown is not directly comparable with OpenHands' internal
timing categories. OpenHands' localization interval includes model-driven
repository exploration. DEIA's $L_0$ interval is a deterministic program
step. Total candidate-generation time is the common timing measure
across the two workflows.

The values are point estimates from one run per task. There are no
repeated runs from which to estimate run-to-run variance. Model scheduling,
caching, and service latency may affect elapsed time. The results support
lower measured resource use for DEIA on this cohort, not a universal
latency guarantee or a reduction in total device-validated repair time.

\begin{rqanswer}{Answer to RQ3.}
On the 29-task cohort, DEIA averages 36,846 tokens, 2.93 model calls, and
25.92 seconds per candidate-generation task. OpenHands averages 199,437
tokens, 7.48 calls, and 39.82 seconds. The difference is primarily in
input-token consumption. These are single-run, generation-stage
measurements and do not include device validation.
\end{rqanswer}

\begin{implicationbox}
\paragraph{Implication.}
Bounded source evidence is associated with lower model input consumption
in this comparison, but the workflow-level results do not isolate the
contribution of individual design choices. Component ablations are needed
to assess those contributions. Repeated runs are also needed to estimate
run-to-run variation.
\end{implicationbox}

\section{Threats to Validity}
\label{sec:threats}

\subsection{Construct Validity}
Reference-based localization measures agreement with curated patches, not
semantic correctness or the only valid repair. Behavioral evidence remains
incomplete: the 29 adjudications include one compile-gate case and one
stub-assisted replay. The matched replay recorded 87 patched-side reachability and smoke checks
per workflow. The original DEIA assessment executed two of the 58
behavior-specific P2P obligations, both for the same repair pattern.
Comprehensive behavior preservation is therefore unestablished. Some expected
outcomes rely on curated references or researcher judgment, creating possible
oracle bias.

\subsection{Internal Validity}
OpenHands lacked a pinned upstream commit, limiting reproduction.
Sequential replay and package adjustments may affect outcomes despite the
shared emulator and harness. Each task had one generation run, and prompt
development overlapped with evaluated or related cases. One mismatched
crash-report task was excluded from matched adjudication but retained in
generation statistics.

\subsection{External Validity}
The evaluation covers 30 ArkTS repair tasks from 24 HarmonyOS applications.
Six reports are synthetic or augmented. All runs used GPT-6.1-Sol and
validation used the HarmonyOS toolchain with an API-26 emulator. Generalization to other
platforms, languages, model providers, and runtime configurations remains
untested.

\FloatBarrier
\section{Related Work}
\label{sec:related}

\subsection{Automated Program Repair}

Automated program repair (APR) investigates how to generate source-code
modifications that address software defects while satisfying specified
correctness criteria. Existing approaches differ in how they identify
faults, construct candidate patches, and assess repair outcomes. Research
has progressed from search-based and pattern-based transformations to
neural repair techniques, with increasing attention to the reliability
of generated patches.

\paragraph{General Program Repair.}
Traditional APR techniques employ different strategies for constructing
candidate patches. GenProg searches program variants through automated
transformations, while PAR learns repair patterns from human-written
patches and TBar applies predefined repair templates
~\cite{legoues2012genprog,kim2013par,liu2019tbar}. Surveys organize
these techniques around fault localization, candidate generation, and
patch validation~\cite{gazzola2019survey}. Neural approaches further
extend repair capabilities across programming languages. For example,
RING combines fault localization, program transformation, and candidate
ranking to support multilingual program repair~\cite{joshi2023ring}.
Defects4J provides a widely adopted benchmark for evaluating APR
techniques on real-world Java defects~\cite{just2014defects4j}. Together,
these approaches establish the foundations of automated patch generation
and evaluation across different repair strategies.

Mobile applications introduce additional execution requirements because
a source-level modification must ultimately be assessed within the target
application runtime. Successful patch application alone does not establish
that an application can launch, reach the relevant interaction, or
preserve its existing behavior on a device. Platform-specific repair also
has to account for build, signing, and runtime constraints.

\paragraph{Mobile Crash Repair.}
Existing research addresses these requirements through crash reproduction,
GUI exploration, and application-level repair. CrashScope automatically
discovers, reports, and reproduces crashes in Android applications
~\cite{moran2016crashscope}. DroidBot generates UI-guided test inputs based
on application states~\cite{li2017droidbot}. Stoat uses guided stochastic
model-based exploration to exercise GUI behaviors~\cite{su2017stoat}.
Sapienz formulates Android application testing as a multi-objective search
problem to expose crashes and other failures~\cite{mao2016sapienz}.
Beyond crash discovery, Droix repairs Android crashes with reusable
transformation operators derived from failing executions
~\cite{tan2018droix}. LLM-CompDroid combines language models with dynamic
analysis to address configuration-compatibility defects in Android
applications~\cite{liu2026compdroid}. These approaches demonstrate the
importance of connecting source-code modifications with application
execution and runtime observations.

\paragraph{HarmonyOS Repair.}
Platform-specific repair has also received increasing attention in the
HarmonyOS ecosystem. Huawei provides official documentation covering
ArkTS language features, static checks, and runtime crash diagnostics
~\cite{huawei2026arkts,huawei2026jscrash}. ArkEval introduces a benchmark
of 502 reproducible ArkTS repair issues and distinguishes build validity
from behavioral repair~\cite{xie2026arkeval}. AppEval further proposes
a unified mobile application repair benchmark spanning ArkTS, Swift, and
Kotlin. Its evaluation contract distinguishes target-program assertions
from infrastructure failures. However, its reported quantitative audit
is Android-specific~\cite{xie2026appeval}. These resources establish
that ArkTS repair benchmarks and platform-oriented evaluation are already
active research topics. They also highlight the need to distinguish
successful patch construction and build execution from evidence of actual
behavioral repair on the target platform.

Against this background, DEIA focuses on maintaining traceable links
among the crash report, source revision, patch artifact, and device
assessment in HarmonyOS crash repair.

\subsection{LLM-Based Repair Agents}

Large language models have extended automated program repair from
individual patch-generation techniques toward interactive workflows
involving repository exploration, tool execution, and iterative feedback.
Existing systems differ in their degree of agent autonomy, the program
information available during repair, and the way execution feedback is
incorporated into subsequent actions. In parallel, repository-level
benchmarks provide increasingly realistic settings for evaluating these
workflows.

\paragraph{Agent-Based Repair.}
Recent surveys classify LLM-based APR systems according to their
model-use patterns, interaction strategies, and repair workflows
~\cite{yang2025survey}. ChatRepair employs conversational interaction
to refine candidate patches, while self-debugging methods use model
feedback and execution results to improve generated programs
~\cite{xia2024chatrepair,chen2024selfdebug}. More autonomous systems
integrate planning and tool execution into the repair process.
RepairAgent coordinates information gathering, patch generation, and
validation through an LLM-based agent~\cite{bouzenia2025repairagent}.
AutoCodeRover combines program-structure information, search, and fault
localization to support repository-level program improvement
~\cite{zhang2024autocoderover}. SWE-agent studies agent--computer
interaction for software engineering~\cite{yang2024sweagent}. OpenHands
provides a general-purpose platform for tool-using software agents
~\cite{wang2024openhands}. Agentless explores a more constrained repair
workflow that organizes software-engineering actions without relying on
an unrestricted autonomous-agent loop~\cite{xia2024agentless}. These
systems demonstrate that repository exploration, tool use, and iterative
validation can be organized through different agent architectures and
levels of autonomy.

\paragraph{Repository-Level Benchmarks.}
The evaluation of LLM-based repair agents has increasingly moved beyond
isolated code snippets toward tasks involving real software repositories.
SWE-bench evaluates whether language models can resolve issues collected
from real-world GitHub repositories~\cite{jimenez2024swebench}.
Subsequent benchmarks extend the task scope to visual software domains
and longer-horizon engineering activities. SWE-bench Multimodal
incorporates visual and user-facing JavaScript tasks, while SWE-Bench Pro
targets longer-horizon, enterprise-oriented repository work
~\cite{yang2025swemultimodal,deng2025swebenchpro}. SWE-Gym provides
executable environments for training software-engineering agents and
verifiers~\cite{pan2025swegym}. SWE-smith focuses on scaling the
construction of software-engineering training tasks
~\cite{yang2025swesmith}. DeepSWE further studies long-horizon engineering
tasks with program-based verifiers~\cite{huang2026deepswe}. Beyond task
completion, studies of agent trajectories and code-execution cost examine
the interactions between model reasoning, tool invocation, and execution
resources~\cite{bouzenia2025trajectories,lin2026executioncost}.

DEIA uses bounded source evidence and content-anchored edits, followed
by deterministic patch materialization and stage-level records linking
generation to validation.

\subsection{Behavioral Validation}

Generating an applicable patch or removing an observed failure does not
necessarily establish that the modified program behaves correctly.
Behavioral validation requires determining which properties should hold
after a modification and whether the available execution evidence is
sufficient to assess them. Existing research addresses this problem
through test-oracle construction, patch assessment, and test generation.

\paragraph{Test Oracles.}
The oracle problem in software testing concerns how to determine whether
an observed execution result is correct when an executable test does not
provide an expected answer~\cite{barr2015oracle}. This problem is
particularly important in automated repair, where a candidate patch may
satisfy the available test suite without preserving the intended program
semantics. Prior studies investigate patch overfitting and develop
techniques for assessing patch correctness beyond test-suite acceptance
~\cite{smith2015overfitting,ye2021patchassessment,lecong2023invalidator}.
Test-generation research further examines how additional inputs and
assertions can improve fault detection and support program repair
~\cite{yu2017testgen}. However, increasing test coverage or the number of
assertions does not automatically establish a trustworthy semantic
oracle. The expected behavior against which an execution is judged
remains a separate requirement.

\paragraph{F2P/P2P Validation.}
A complementary distinction in repair evaluation concerns the different
behavioral objectives of fail-to-pass (F2P) and pass-to-pass (P2P) checks.
F2P checks target the removal of an existing failure, whereas P2P checks
examine whether previously valid behavior is preserved after a
modification. These distinctions already appear in repository-level
repair benchmarks and platform-oriented evaluation protocols
~\cite{jimenez2024swebench,xie2026appeval}. The categories themselves
are not novel. Their effectiveness nevertheless depends on the quality
of the expected-behavior specifications and the reliability of the
corresponding execution oracles. A test that can be executed but lacks a
supported expected outcome cannot provide sufficient evidence of
behavioral correctness. Similarly, a test whose required interaction or
observable signal is unavailable cannot be treated as a completed
validation.

Building on established F2P/P2P and test-oracle concepts, DEIA represents
each fault- or change-derived check as an obligation with an expected
property, evidence source, prerequisites, and observable signal. Checks
lacking sufficient support or observability remain pending.

\section{Conclusion and Future Work}
\label{sec:conclusion}

DEIA links each candidate patch to the crash report that motivated it, the
source it changes, and the build and replay records used to assess it.
It specifies which behaviors need checking and which observations would
support each judgment. Across 30 tasks, DEIA produced 30 candidate patches.
All 30 generated patches passed the applicability check. The 30 delivered
patches were byte-identical to the generated artifacts and also passed
the check. On 29 matched tasks, the first added line exactly matched a
reference edit location in 25 DEIA patches and 19 OpenHands patches.
With one run per workflow and task, DEIA used 81.5\% fewer tokens, made
60.8\\
measurements end at patch materialization.

Device assessment covered 28 runtime fault cases and one build-gate case.
Each workflow also had 87 patched-side reachability and smoke observations.
The original DEIA assessment executed two of the 58 behavior-specific P2P
checks, both for the same repair pattern. The remaining checks and untested
paths leave the broader behavioral effects of these patches open.

Future work will focus on verifying behavior that the current replay
leaves unresolved. One priority is to run the same valid interaction on the
original and patched applications and check both outcomes against an
independently reviewed expectation. A regression control would then
introduce a patch that removes the crash while breaking that interaction.
For example, a list might stop crashing at its boundary but silently ignore
a valid reorder. Detecting that change on the emulator would test whether
the P2P check can expose a regression even when the F2P check passes.
Repeated runs on fresh crash cases with a pinned baseline would separately
quantify variation in repair outcomes and generation cost.

\bibliographystyle{ACM-Reference-Format}
\bibliography{refs}


\begin{thebibliography}{37}


\ifx \showCODEN    \undefined \def \showCODEN     #1{\unskip}     \fi
\ifx \showDOI      \undefined \def \showDOI       #1{#1}\fi
\ifx \showISBNx    \undefined \def \showISBNx     #1{\unskip}     \fi
\ifx \showISBNxiii \undefined \def \showISBNxiii  #1{\unskip}     \fi
\ifx \showISSN     \undefined \def \showISSN      #1{\unskip}     \fi
\ifx \showLCCN     \undefined \def \showLCCN      #1{\unskip}     \fi
\ifx \shownote     \undefined \def \shownote      #1{#1}          \fi
\ifx \showarticletitle \undefined \def \showarticletitle #1{#1}   \fi
\ifx \showURL      \undefined \def \showURL       {\relax}        \fi
\providecommand\bibfield[2]{#2}
\providecommand\bibinfo[2]{#2}
\providecommand\natexlab[1]{#1}
\providecommand\showeprint[2][]{arXiv:#2}

\bibitem[Barr et~al\mbox{.}(2015)]%
        {barr2015oracle}
\bibfield{author}{\bibinfo{person}{Earl~T. Barr}, \bibinfo{person}{Mark
  Harman}, \bibinfo{person}{Phil McMinn}, \bibinfo{person}{Muzammil Shahbaz},
  {and} \bibinfo{person}{Shin Yoo}.} \bibinfo{year}{2015}\natexlab{}.
\newblock \showarticletitle{The Oracle Problem in Software Testing: A Survey}.
\newblock \bibinfo{journal}{\emph{IEEE Transactions on Software Engineering}}
  \bibinfo{volume}{41}, \bibinfo{number}{5} (\bibinfo{year}{2015}),
  \bibinfo{pages}{507--525}.
\newblock
\urldef\tempurl%
\url{https://doi.org/10.1109/TSE.2014.2372785}
\showDOI{\tempurl}


\bibitem[Bouzenia et~al\mbox{.}(2025)]%
        {bouzenia2025repairagent}
\bibfield{author}{\bibinfo{person}{Islem Bouzenia}, \bibinfo{person}{Premkumar
  Devanbu}, {and} \bibinfo{person}{Michael Pradel}.}
  \bibinfo{year}{2025}\natexlab{}.
\newblock \showarticletitle{{RepairAgent}: An Autonomous, {LLM}-Based Agent for
  Program Repair}. In \bibinfo{booktitle}{\emph{Proceedings of the 47th
  IEEE/ACM International Conference on Software Engineering (ICSE)}}.
  \bibinfo{publisher}{IEEE}, \bibinfo{address}{Ottawa, Canada},
  \bibinfo{pages}{2188--2200}.
\newblock
\urldef\tempurl%
\url{https://doi.org/10.1109/ICSE55347.2025.00157}
\showDOI{\tempurl}


\bibitem[Bouzenia and Pradel(2025)]%
        {bouzenia2025trajectories}
\bibfield{author}{\bibinfo{person}{Islem Bouzenia} {and}
  \bibinfo{person}{Michael Pradel}.} \bibinfo{year}{2025}\natexlab{}.
\newblock \showarticletitle{Understanding Software Engineering Agents: A Study
  of Thought-Action-Result Trajectories}. In
  \bibinfo{booktitle}{\emph{Proceedings of the 40th IEEE/ACM International
  Conference on Automated Software Engineering (ASE)}}.
  \bibinfo{publisher}{IEEE}, \bibinfo{address}{Seoul, Korea}.
\newblock
\urldef\tempurl%
\url{https://doi.org/10.1109/ASE63991.2025.00234}
\showDOI{\tempurl}


\bibitem[Chen et~al\mbox{.}(2024)]%
        {chen2024selfdebug}
\bibfield{author}{\bibinfo{person}{Xinyun Chen}, \bibinfo{person}{Maxwell Lin},
  \bibinfo{person}{Nathanael Sch{\"a}rli}, {and} \bibinfo{person}{Denny Zhou}.}
  \bibinfo{year}{2024}\natexlab{}.
\newblock \showarticletitle{Teaching Large Language Models to Self-Debug}. In
  \bibinfo{booktitle}{\emph{International Conference on Learning
  Representations (ICLR)}}. \bibinfo{publisher}{OpenReview.net},
  \bibinfo{address}{Vienna, Austria}.
\newblock


\bibitem[Deng et~al\mbox{.}(2025)]%
        {deng2025swebenchpro}
\bibfield{author}{\bibinfo{person}{Xiang Deng}, \bibinfo{person}{Jeff Da},
  \bibinfo{person}{Edwin Pan}, \bibinfo{person}{Yannis~Yiming He},
  \bibinfo{person}{Charles Ide}, \bibinfo{person}{Kanak Garg},
  \bibinfo{person}{Niklas Lauffer}, \bibinfo{person}{Andrew Park},
  \bibinfo{person}{Nitin Pasari}, \bibinfo{person}{Chetan Rane},
  \bibinfo{person}{Karmini Sampath}, \bibinfo{person}{Maya Krishnan},
  \bibinfo{person}{Srivatsa Kundurthy}, \bibinfo{person}{Sean Hendryx},
  \bibinfo{person}{Zifan Wang}, \bibinfo{person}{Chen Bo~Calvin Zhang},
  \bibinfo{person}{Noah Jacobson}, \bibinfo{person}{Bing Liu}, {and}
  \bibinfo{person}{Brad Kenstler}.} \bibinfo{year}{2025}\natexlab{}.
\newblock \bibinfo{title}{{SWE-Bench Pro}: Can {AI} Agents Solve Long-Horizon
  Software Engineering Tasks?}
\newblock
\newblock
\showeprint[arxiv]{2509.16941}~[cs.SE]
\urldef\tempurl%
\url{https://arxiv.org/abs/2509.16941}
\showURL{%
\tempurl}


\bibitem[Gazzola et~al\mbox{.}(2019)]%
        {gazzola2019survey}
\bibfield{author}{\bibinfo{person}{Luca Gazzola}, \bibinfo{person}{Daniela
  Micucci}, {and} \bibinfo{person}{Leonardo Mariani}.}
  \bibinfo{year}{2019}\natexlab{}.
\newblock \showarticletitle{Automatic Software Repair: A Survey}.
\newblock \bibinfo{journal}{\emph{IEEE Transactions on Software Engineering}}
  \bibinfo{volume}{45}, \bibinfo{number}{1} (\bibinfo{year}{2019}),
  \bibinfo{pages}{34--67}.
\newblock
\urldef\tempurl%
\url{https://doi.org/10.1109/TSE.2017.2755013}
\showDOI{\tempurl}


\bibitem[Goues et~al\mbox{.}(2012)]%
        {legoues2012genprog}
\bibfield{author}{\bibinfo{person}{Claire~Le Goues}, \bibinfo{person}{ThanhVu
  Nguyen}, \bibinfo{person}{Stephanie Forrest}, {and} \bibinfo{person}{Westley
  Weimer}.} \bibinfo{year}{2012}\natexlab{}.
\newblock \showarticletitle{{GenProg}: A Generic Method for Automatic Software
  Repair}.
\newblock \bibinfo{journal}{\emph{IEEE Transactions on Software Engineering}}
  \bibinfo{volume}{38}, \bibinfo{number}{1} (\bibinfo{year}{2012}),
  \bibinfo{pages}{54--72}.
\newblock
\urldef\tempurl%
\url{https://doi.org/10.1109/TSE.2011.104}
\showDOI{\tempurl}


\bibitem[Huang et~al\mbox{.}(2026)]%
        {huang2026deepswe}
\bibfield{author}{\bibinfo{person}{Wenqi Huang}, \bibinfo{person}{Charley Lee},
  \bibinfo{person}{Leonard Tng}, {and} \bibinfo{person}{Serena Ge}.}
  \bibinfo{year}{2026}\natexlab{}.
\newblock \bibinfo{title}{{DeepSWE}: Measuring Frontier Coding Agents on
  Original, Long-Horizon Engineering Tasks}.
\newblock
\newblock
\showeprint[arxiv]{2607.07946}~[cs.SE]
\urldef\tempurl%
\url{https://arxiv.org/abs/2607.07946}
\showURL{%
\tempurl}


\bibitem[{Huawei}(2026a)]%
        {huawei2026arkts}
\bibfield{author}{\bibinfo{person}{{Huawei}}.}
  \bibinfo{year}{2026}\natexlab{a}.
\newblock \bibinfo{title}{{ArkTS}: Introduction}.
\newblock
\newblock
\urldef\tempurl%
\url{https://developer.huawei.com/consumer/en/arkts/}
\showURL{%
\tempurl}
\newblock
\shownote{Official HarmonyOS developer documentation; accessed 2026-11-07}.


\bibitem[{Huawei}(2026b)]%
        {huawei2026jscrash}
\bibfield{author}{\bibinfo{person}{{Huawei}}.}
  \bibinfo{year}{2026}\natexlab{b}.
\newblock \bibinfo{title}{{JS Crash} (Process Crash) Detection}.
\newblock
\newblock
\urldef\tempurl%
\url{https://developer.huawei.com/consumer/en/doc/harmonyos-guides/jscrash-guidelines/}
\showURL{%
\tempurl}
\newblock
\shownote{Official HarmonyOS developer documentation; accessed 2026-11-07}.


\bibitem[Jimenez et~al\mbox{.}(2024)]%
        {jimenez2024swebench}
\bibfield{author}{\bibinfo{person}{Carlos~E. Jimenez}, \bibinfo{person}{John
  Yang}, \bibinfo{person}{Alexander Wettig}, \bibinfo{person}{Shunyu Yao},
  \bibinfo{person}{Kexin Pei}, \bibinfo{person}{Ofir Press}, {and}
  \bibinfo{person}{Karthik Narasimhan}.} \bibinfo{year}{2024}\natexlab{}.
\newblock \showarticletitle{{SWE-bench}: Can Language Models Resolve Real-World
  {GitHub} Issues?}. In \bibinfo{booktitle}{\emph{International Conference on
  Learning Representations (ICLR)}}. \bibinfo{publisher}{OpenReview.net},
  \bibinfo{address}{Vienna, Austria}.
\newblock


\bibitem[Joshi et~al\mbox{.}(2023)]%
        {joshi2023ring}
\bibfield{author}{\bibinfo{person}{Harshit Joshi}, \bibinfo{person}{Jos{\'e}
  Cambronero}, \bibinfo{person}{Sumit Gulwani}, \bibinfo{person}{Vu Le},
  \bibinfo{person}{Ivan Radicek}, {and} \bibinfo{person}{Gust Verbruggen}.}
  \bibinfo{year}{2023}\natexlab{}.
\newblock \showarticletitle{Repair Is Nearly Generation: Multilingual Program
  Repair with {LLMs}}. In \bibinfo{booktitle}{\emph{Proceedings of the AAAI
  Conference on Artificial Intelligence}}, Vol.~\bibinfo{volume}{37}.
  \bibinfo{publisher}{AAAI Press}, \bibinfo{address}{Washington, DC, USA},
  \bibinfo{pages}{5131--5140}.
\newblock
\urldef\tempurl%
\url{https://doi.org/10.1609/aaai.v37i4.25642}
\showDOI{\tempurl}


\bibitem[Just et~al\mbox{.}(2014)]%
        {just2014defects4j}
\bibfield{author}{\bibinfo{person}{Ren{\'e} Just}, \bibinfo{person}{Darioush
  Jalali}, {and} \bibinfo{person}{Michael~D. Ernst}.}
  \bibinfo{year}{2014}\natexlab{}.
\newblock \showarticletitle{{Defects4J}: A Database of Existing Faults to
  Enable Controlled Testing Studies for Java Programs}. In
  \bibinfo{booktitle}{\emph{Proceedings of the 23rd International Symposium on
  Software Testing and Analysis (ISSTA)}}. \bibinfo{publisher}{ACM},
  \bibinfo{address}{San Jose, CA, USA}, \bibinfo{pages}{437--440}.
\newblock
\urldef\tempurl%
\url{https://doi.org/10.1145/2610384.2628055}
\showDOI{\tempurl}


\bibitem[Kim et~al\mbox{.}(2013)]%
        {kim2013par}
\bibfield{author}{\bibinfo{person}{Dongsun Kim}, \bibinfo{person}{Jaechang
  Nam}, \bibinfo{person}{Jaewoo Song}, {and} \bibinfo{person}{Sunghun Kim}.}
  \bibinfo{year}{2013}\natexlab{}.
\newblock \showarticletitle{Automatic Patch Generation Learned from
  Human-Written Patches}. In \bibinfo{booktitle}{\emph{Proceedings of the 35th
  International Conference on Software Engineering (ICSE)}}.
  \bibinfo{publisher}{IEEE}, \bibinfo{address}{San Francisco, CA, USA},
  \bibinfo{pages}{802--811}.
\newblock
\urldef\tempurl%
\url{https://doi.org/10.1109/ICSE.2013.6606626}
\showDOI{\tempurl}


\bibitem[Le-Cong et~al\mbox{.}(2023)]%
        {lecong2023invalidator}
\bibfield{author}{\bibinfo{person}{Thanh Le-Cong}, \bibinfo{person}{Duc-Minh
  Luong}, \bibinfo{person}{Xuan Bach~D. Le}, \bibinfo{person}{David Lo},
  \bibinfo{person}{Nhat-Hoa Tran}, {and} \bibinfo{person}{Quyet-Thang Huynh}.}
  \bibinfo{year}{2023}\natexlab{}.
\newblock \bibinfo{title}{{Invalidator}: Automated Patch Correctness Assessment
  via Semantic and Syntactic Reasoning}.
\newblock
\newblock
\showeprint[arxiv]{2301.01113}


\bibitem[Li et~al\mbox{.}(2017)]%
        {li2017droidbot}
\bibfield{author}{\bibinfo{person}{Yuanchun Li}, \bibinfo{person}{Ziyue Yang},
  \bibinfo{person}{Yao Guo}, {and} \bibinfo{person}{Xiangqun Chen}.}
  \bibinfo{year}{2017}\natexlab{}.
\newblock \showarticletitle{{DroidBot}: A Lightweight {UI}-Guided Test Input
  Generator for {Android}}. In \bibinfo{booktitle}{\emph{2017 IEEE/ACM 39th
  IEEE International Conference on Software Engineering Companion (ICSE-C)}}.
  \bibinfo{publisher}{IEEE}, \bibinfo{address}{Buenos Aires, Argentina},
  \bibinfo{pages}{23--26}.
\newblock
\urldef\tempurl%
\url{https://doi.org/10.1109/ICSE-C.2017.8}
\showDOI{\tempurl}


\bibitem[Lin et~al\mbox{.}(2026)]%
        {lin2026executioncost}
\bibfield{author}{\bibinfo{person}{Zhihao Lin}, \bibinfo{person}{Junhua Zhu},
  \bibinfo{person}{Mingyi Zhou}, \bibinfo{person}{Xin Wang},
  \bibinfo{person}{Zhensu Sun}, \bibinfo{person}{Renyu Yang},
  \bibinfo{person}{David Lo}, {and} \bibinfo{person}{Li Li}.}
  \bibinfo{year}{2026}\natexlab{}.
\newblock \bibinfo{title}{To Run or Not to Run: Analyzing the
  Cost-Effectiveness of Code Execution in {LLM}-Based Program Repair}.
\newblock
\newblock
\showeprint[arxiv]{2606.26978}~[cs.SE]
\urldef\tempurl%
\url{https://arxiv.org/abs/2606.26978}
\showURL{%
\tempurl}


\bibitem[Liu et~al\mbox{.}(2019)]%
        {liu2019tbar}
\bibfield{author}{\bibinfo{person}{Kui Liu}, \bibinfo{person}{Anil Koyuncu},
  \bibinfo{person}{Dongsun Kim}, {and} \bibinfo{person}{Tegawend{\'e}~F.
  Bissyand{\'e}}.} \bibinfo{year}{2019}\natexlab{}.
\newblock \showarticletitle{{TBar}: Revisiting Template-Based Automated Program
  Repair}. In \bibinfo{booktitle}{\emph{Proceedings of the 28th ACM SIGSOFT
  International Symposium on Software Testing and Analysis (ISSTA)}}.
  \bibinfo{publisher}{ACM}, \bibinfo{address}{Beijing, China},
  \bibinfo{pages}{31--41}.
\newblock
\urldef\tempurl%
\url{https://doi.org/10.1145/3293882.3330577}
\showDOI{\tempurl}


\bibitem[Liu et~al\mbox{.}(2026)]%
        {liu2026compdroid}
\bibfield{author}{\bibinfo{person}{Zhijie Liu}, \bibinfo{person}{Yutian Tang},
  \bibinfo{person}{Meiyun Li}, \bibinfo{person}{Xin Jin},
  \bibinfo{person}{Yunfei Long}, \bibinfo{person}{Liang~Feng Zhang}, {and}
  \bibinfo{person}{Xiapu Luo}.} \bibinfo{year}{2026}\natexlab{}.
\newblock \showarticletitle{{LLM-CompDroid}: Repairing Configuration
  Compatibility Bugs in Android Apps with Pre-trained Large Language Models}.
\newblock \bibinfo{journal}{\emph{ACM Transactions on Software Engineering and
  Methodology}} \bibinfo{volume}{35}, \bibinfo{number}{3}
  (\bibinfo{year}{2026}), \bibinfo{pages}{63}.
\newblock
\urldef\tempurl%
\url{https://doi.org/10.1145/3736406}
\showDOI{\tempurl}


\bibitem[Mao et~al\mbox{.}(2016)]%
        {mao2016sapienz}
\bibfield{author}{\bibinfo{person}{Ke Mao}, \bibinfo{person}{Mark Harman},
  {and} \bibinfo{person}{Yue Jia}.} \bibinfo{year}{2016}\natexlab{}.
\newblock \showarticletitle{{Sapienz}: Multi-Objective Automated Testing for
  Android Applications}. In \bibinfo{booktitle}{\emph{Proceedings of the 25th
  ACM SIGSOFT International Symposium on Software Testing and Analysis
  (ISSTA)}}. \bibinfo{publisher}{ACM}, \bibinfo{address}{Saarbr{\"u}cken,
  Germany}, \bibinfo{pages}{94--105}.
\newblock
\urldef\tempurl%
\url{https://doi.org/10.1145/2931037.2931054}
\showDOI{\tempurl}


\bibitem[Moran et~al\mbox{.}(2016)]%
        {moran2016crashscope}
\bibfield{author}{\bibinfo{person}{Kevin Moran}, \bibinfo{person}{Mario
  Linares-V{\'a}squez}, \bibinfo{person}{Carlos Bernal-C{\'a}rdenas},
  \bibinfo{person}{Christopher Vendome}, {and} \bibinfo{person}{Denys
  Poshyvanyk}.} \bibinfo{year}{2016}\natexlab{}.
\newblock \showarticletitle{Automatically Discovering, Reporting and
  Reproducing {Android} Application Crashes}. In \bibinfo{booktitle}{\emph{2016
  IEEE International Conference on Software Testing, Verification and
  Validation (ICST)}}. \bibinfo{publisher}{IEEE}, \bibinfo{address}{Chicago,
  IL, USA}, \bibinfo{pages}{33--44}.
\newblock
\urldef\tempurl%
\url{https://doi.org/10.1109/ICST.2016.34}
\showDOI{\tempurl}


\bibitem[Pan et~al\mbox{.}(2025)]%
        {pan2025swegym}
\bibfield{author}{\bibinfo{person}{Jiayi Pan}, \bibinfo{person}{Xingyao Wang},
  \bibinfo{person}{Graham Neubig}, \bibinfo{person}{Navdeep Jaitly},
  \bibinfo{person}{Heng Ji}, \bibinfo{person}{Alane Suhr}, {and}
  \bibinfo{person}{Yizhe Zhang}.} \bibinfo{year}{2025}\natexlab{}.
\newblock \showarticletitle{Training Software Engineering Agents and Verifiers
  with {SWE-Gym}}. In \bibinfo{booktitle}{\emph{Proceedings of the 42nd
  International Conference on Machine Learning}}
  \emph{(\bibinfo{series}{Proceedings of Machine Learning Research},
  Vol.~\bibinfo{volume}{267})}. \bibinfo{publisher}{PMLR},
  \bibinfo{address}{Vancouver, Canada}, \bibinfo{pages}{47717--47737}.
\newblock
\urldef\tempurl%
\url{https://proceedings.mlr.press/v267/pan25g.html}
\showURL{%
\tempurl}


\bibitem[Smith et~al\mbox{.}(2015)]%
        {smith2015overfitting}
\bibfield{author}{\bibinfo{person}{Edward~K. Smith}, \bibinfo{person}{Earl~T.
  Barr}, \bibinfo{person}{Claire~Le Goues}, {and} \bibinfo{person}{Yuriy
  Brun}.} \bibinfo{year}{2015}\natexlab{}.
\newblock \showarticletitle{Is the Cure Worse Than the Disease? Overfitting in
  Automated Program Repair}. In \bibinfo{booktitle}{\emph{Proceedings of the
  2015 10th Joint Meeting on Foundations of Software Engineering (ESEC/FSE)}}.
  \bibinfo{publisher}{ACM}, \bibinfo{address}{Bergamo, Italy},
  \bibinfo{pages}{532--543}.
\newblock


\bibitem[Su et~al\mbox{.}(2017)]%
        {su2017stoat}
\bibfield{author}{\bibinfo{person}{Ting Su}, \bibinfo{person}{Guozhu Meng},
  \bibinfo{person}{Yuting Chen}, \bibinfo{person}{Ke Wu},
  \bibinfo{person}{Weiming Yang}, \bibinfo{person}{Yao Yao},
  \bibinfo{person}{Geguang Pu}, \bibinfo{person}{Yang Liu}, {and}
  \bibinfo{person}{Zhendong Su}.} \bibinfo{year}{2017}\natexlab{}.
\newblock \showarticletitle{Guided, Stochastic Model-Based {GUI} Testing of
  {Android} Apps}. In \bibinfo{booktitle}{\emph{Proceedings of the 2017 ACM
  Joint European Software Engineering Conference and Symposium on the
  Foundations of Software Engineering (ESEC/FSE)}}. \bibinfo{publisher}{ACM},
  \bibinfo{address}{Paderborn, Germany}, \bibinfo{pages}{245--256}.
\newblock
\urldef\tempurl%
\url{https://doi.org/10.1145/3106237.3106298}
\showDOI{\tempurl}


\bibitem[Tan et~al\mbox{.}(2018)]%
        {tan2018droix}
\bibfield{author}{\bibinfo{person}{Shin~Hwei Tan}, \bibinfo{person}{Zhen Dong},
  \bibinfo{person}{Xiang Gao}, {and} \bibinfo{person}{Abhik Roychoudhury}.}
  \bibinfo{year}{2018}\natexlab{}.
\newblock \showarticletitle{Repairing Crashes in {Android} Apps}. In
  \bibinfo{booktitle}{\emph{Proceedings of the 40th International Conference on
  Software Engineering (ICSE)}}. \bibinfo{publisher}{ACM},
  \bibinfo{address}{Gothenburg, Sweden}, \bibinfo{pages}{187--198}.
\newblock
\urldef\tempurl%
\url{https://doi.org/10.1145/3180155.3180243}
\showDOI{\tempurl}


\bibitem[Wang et~al\mbox{.}(2025)]%
        {wang2024openhands}
\bibfield{author}{\bibinfo{person}{Xingyao Wang}, \bibinfo{person}{Boxuan Li},
  \bibinfo{person}{Yufan Song}, \bibinfo{person}{Frank~F. Xu},
  \bibinfo{person}{Xiangru Tang}, \bibinfo{person}{Mingchen Zhuge},
  \bibinfo{person}{Jiayi Pan}, \bibinfo{person}{Yueqi Song},
  \bibinfo{person}{Bowen Li}, \bibinfo{person}{Jaskirat Singh},
  \bibinfo{person}{Hoang~H. Tran}, \bibinfo{person}{Fuqiang Li},
  \bibinfo{person}{Ren Ma}, \bibinfo{person}{Mingzhang Zheng},
  \bibinfo{person}{Bill Qian}, \bibinfo{person}{Daniel Shao},
  \bibinfo{person}{Niklas Muennighoff}, \bibinfo{person}{Yizhe Zhang},
  \bibinfo{person}{Binyuan Hui}, \bibinfo{person}{Junyang Lin},
  \bibinfo{person}{Robert Brennan}, \bibinfo{person}{Hao Peng},
  \bibinfo{person}{Heng Ji}, {and} \bibinfo{person}{Graham Neubig}.}
  \bibinfo{year}{2025}\natexlab{}.
\newblock \showarticletitle{{OpenHands}: An Open Platform for {AI} Software
  Developers as Generalist Agents}. In \bibinfo{booktitle}{\emph{International
  Conference on Learning Representations (ICLR)}}.
  \bibinfo{publisher}{OpenReview.net}, \bibinfo{address}{Singapore}.
\newblock
\urldef\tempurl%
\url{https://arxiv.org/abs/2407.16741}
\showURL{%
\tempurl}


\bibitem[Xia et~al\mbox{.}(2024a)]%
        {xia2024agentless}
\bibfield{author}{\bibinfo{person}{Chunqiu~Steven Xia}, \bibinfo{person}{Yinlin
  Deng}, \bibinfo{person}{Soren Dunn}, {and} \bibinfo{person}{Lingming Zhang}.}
  \bibinfo{year}{2024}\natexlab{a}.
\newblock \bibinfo{title}{{Agentless}: Demystifying LLM-based Software
  Engineering Agents}.
\newblock
\newblock
\showeprint[arxiv]{2407.01489}


\bibitem[Xia et~al\mbox{.}(2024b)]%
        {xia2024chatrepair}
\bibfield{author}{\bibinfo{person}{Chunqiu~Steven Xia},
  \bibinfo{person}{Yuxiang Wei}, {and} \bibinfo{person}{Lingming Zhang}.}
  \bibinfo{year}{2024}\natexlab{b}.
\newblock \showarticletitle{Automated Program Repair via Conversation: Fixing
  162 out of 337 Bugs for {\$}0.42 Each Using {ChatGPT}}. In
  \bibinfo{booktitle}{\emph{Proceedings of the 33rd ACM SIGSOFT International
  Symposium on Software Testing and Analysis (ISSTA)}}.
  \bibinfo{publisher}{ACM}, \bibinfo{address}{Vienna, Austria},
  \bibinfo{pages}{819--831}.
\newblock
\urldef\tempurl%
\url{https://doi.org/10.1145/3650212.3680323}
\showDOI{\tempurl}


\bibitem[Xie et~al\mbox{.}(2026a)]%
        {xie2026appeval}
\bibfield{author}{\bibinfo{person}{Bang Xie}, \bibinfo{person}{Hao Liu},
  \bibinfo{person}{Zhenyu Shi}, \bibinfo{person}{Yonghao Zhang},
  \bibinfo{person}{Senjian Zhang}, \bibinfo{person}{Zhiyuan Peng},
  \bibinfo{person}{Xin Yin}, \bibinfo{person}{Chenhao Ying},
  \bibinfo{person}{Yuan Luo}, \bibinfo{person}{Wei Chen},
  \bibinfo{person}{Haiming Jin}, \bibinfo{person}{Shaocong Long},
  \bibinfo{person}{Xu Liu}, {and} \bibinfo{person}{Zhe Peng}.}
  \bibinfo{year}{2026}\natexlab{a}.
\newblock \bibinfo{title}{{AppEval}: A Unified Benchmark for LLM-Based Mobile
  Application Repair in {ArkTS}, {Swift}, and {Kotlin}}.
\newblock
\newblock
\showeprint[arxiv]{2608.18588}
\urldef\tempurl%
\url{https://arxiv.org/abs/2608.18588}
\showURL{%
\tempurl}


\bibitem[Xie et~al\mbox{.}(2026b)]%
        {xie2026arkeval}
\bibfield{author}{\bibinfo{person}{Bang Xie}, \bibinfo{person}{Senjian Zhang},
  \bibinfo{person}{Zhiyuan Peng}, \bibinfo{person}{Wei Chen},
  \bibinfo{person}{Xin Yin}, \bibinfo{person}{Chenhao Ying}, {and}
  \bibinfo{person}{Yuan Luo}.} \bibinfo{year}{2026}\natexlab{b}.
\newblock \showarticletitle{{ArkEval}: Benchmarking and Evaluating Automated
  Code Repair for {ArkTS}}. In \bibinfo{booktitle}{\emph{Proceedings of the
  41st IEEE/ACM International Conference on Automated Software Engineering
  (ASE)}}. \bibinfo{publisher}{ACM}, \bibinfo{address}{Munich, Germany}.
\newblock
\urldef\tempurl%
\url{https://doi.org/10.1145/3832783.3837514}
\showDOI{\tempurl}


\bibitem[Yang et~al\mbox{.}(2025a)]%
        {yang2025survey}
\bibfield{author}{\bibinfo{person}{Boyang Yang}, \bibinfo{person}{Zijian Cai},
  \bibinfo{person}{Fengling Liu}, \bibinfo{person}{Bach Le},
  \bibinfo{person}{Lingming Zhang}, \bibinfo{person}{Tegawend{\'e}~F.
  Bissyand{\'e}}, \bibinfo{person}{Yang Liu}, {and} \bibinfo{person}{Haoye
  Tian}.} \bibinfo{year}{2025}\natexlab{a}.
\newblock \bibinfo{title}{A Survey of LLM-Based Automated Program Repair:
  Taxonomies, Design Paradigms, and Applications}.
\newblock
\newblock
\showeprint[arxiv]{2506.23749}


\bibitem[Yang et~al\mbox{.}(2024)]%
        {yang2024sweagent}
\bibfield{author}{\bibinfo{person}{John Yang}, \bibinfo{person}{Carlos~E.
  Jimenez}, \bibinfo{person}{Alexander Wettig}, \bibinfo{person}{Kilian
  Lieret}, \bibinfo{person}{Shunyu Yao}, \bibinfo{person}{Karthik Narasimhan},
  {and} \bibinfo{person}{Ofir Press}.} \bibinfo{year}{2024}\natexlab{}.
\newblock \showarticletitle{{SWE-agent}: Agent--Computer Interfaces Enable
  Automated Software Engineering}. In \bibinfo{booktitle}{\emph{Advances in
  Neural Information Processing Systems (NeurIPS)}}. \bibinfo{publisher}{Curran
  Associates, Inc.}, \bibinfo{address}{Vancouver, Canada}.
\newblock
\urldef\tempurl%
\url{https://doi.org/10.52202/079017-1601}
\showDOI{\tempurl}


\bibitem[Yang et~al\mbox{.}(2025b)]%
        {yang2025swemultimodal}
\bibfield{author}{\bibinfo{person}{John Yang}, \bibinfo{person}{Carlos~E.
  Jimenez}, \bibinfo{person}{Alex~L. Zhang}, \bibinfo{person}{Kilian Lieret},
  \bibinfo{person}{Joyce Yang}, \bibinfo{person}{Xindi Wu},
  \bibinfo{person}{Ori Press}, \bibinfo{person}{Niklas Muennighoff},
  \bibinfo{person}{Gabriel Synnaeve}, \bibinfo{person}{Karthik~R. Narasimhan},
  \bibinfo{person}{Diyi Yang}, \bibinfo{person}{Sida~I. Wang}, {and}
  \bibinfo{person}{Ofir Press}.} \bibinfo{year}{2025}\natexlab{b}.
\newblock \showarticletitle{{SWE-bench Multimodal}: Do {AI} Systems Generalize
  to Visual Software Domains?}. In \bibinfo{booktitle}{\emph{International
  Conference on Learning Representations (ICLR)}}.
  \bibinfo{publisher}{OpenReview.net}, \bibinfo{address}{Singapore}.
\newblock
\urldef\tempurl%
\url{https://openreview.net/forum?id=riTiq3i21b}
\showURL{%
\tempurl}


\bibitem[Yang et~al\mbox{.}(2025c)]%
        {yang2025swesmith}
\bibfield{author}{\bibinfo{person}{John Yang}, \bibinfo{person}{Kilian Lieret},
  \bibinfo{person}{Carlos~E. Jimenez}, \bibinfo{person}{Alexander Wettig},
  \bibinfo{person}{Kabir Khandpur}, \bibinfo{person}{Yanzhe Zhang},
  \bibinfo{person}{Binyuan Hui}, \bibinfo{person}{Ofir Press},
  \bibinfo{person}{Ludwig Schmidt}, {and} \bibinfo{person}{Diyi Yang}.}
  \bibinfo{year}{2025}\natexlab{c}.
\newblock \showarticletitle{{SWE-smith}: Scaling Data for Software Engineering
  Agents}. In \bibinfo{booktitle}{\emph{Advances in Neural Information
  Processing Systems (NeurIPS) Datasets and Benchmarks}}.
  \bibinfo{publisher}{Curran Associates, Inc.}, \bibinfo{address}{San Diego,
  CA, USA}.
\newblock
\showeprint[arxiv]{2504.21798}
\urldef\tempurl%
\url{https://arxiv.org/abs/2504.21798}
\showURL{%
\tempurl}


\bibitem[Ye et~al\mbox{.}(2021)]%
        {ye2021patchassessment}
\bibfield{author}{\bibinfo{person}{He Ye}, \bibinfo{person}{Matias Martinez},
  {and} \bibinfo{person}{Martin Monperrus}.} \bibinfo{year}{2021}\natexlab{}.
\newblock \showarticletitle{Automated Patch Assessment for Program Repair at
  Scale}.
\newblock \bibinfo{journal}{\emph{Empirical Software Engineering}}
  \bibinfo{volume}{26}, \bibinfo{number}{2} (\bibinfo{year}{2021}),
  \bibinfo{pages}{20}.
\newblock
\urldef\tempurl%
\url{https://doi.org/10.1007/s10664-020-09920-w}
\showDOI{\tempurl}


\bibitem[Yu et~al\mbox{.}(2017)]%
        {yu2017testgen}
\bibfield{author}{\bibinfo{person}{Zhongxing Yu}, \bibinfo{person}{Matias
  Martinez}, \bibinfo{person}{Benjamin Danglot}, \bibinfo{person}{Thomas
  Durieux}, {and} \bibinfo{person}{Martin Monperrus}.}
  \bibinfo{year}{2017}\natexlab{}.
\newblock \bibinfo{title}{Test Case Generation for Program Repair: A Study of
  Feasibility and Effectiveness}.
\newblock
\newblock
\showeprint[arxiv]{1703.00198}


\bibitem[Zhang et~al\mbox{.}(2024)]%
        {zhang2024autocoderover}
\bibfield{author}{\bibinfo{person}{Yuntong Zhang}, \bibinfo{person}{Haifeng
  Ruan}, \bibinfo{person}{Zhiyu Fan}, {and} \bibinfo{person}{Abhik
  Roychoudhury}.} \bibinfo{year}{2024}\natexlab{}.
\newblock \showarticletitle{{AutoCodeRover}: Autonomous Program Improvement}.
  In \bibinfo{booktitle}{\emph{Proceedings of the 33rd ACM SIGSOFT
  International Symposium on Software Testing and Analysis (ISSTA)}}.
  \bibinfo{publisher}{ACM}, \bibinfo{address}{Vienna, Austria},
  \bibinfo{pages}{1592--1604}.
\newblock
\urldef\tempurl%
\url{https://doi.org/10.1145/3650212.3680384}
\showDOI{\tempurl}


\end{thebibliography}

\end{document}